\documentclass[twocolumn,superscriptaddress,floatfix,
preprintnumbers,prd,nofootinbib,10pt,aps]{revtex4-1}

\usepackage[colorlinks=true,breaklinks=true]{hyperref}
\hypersetup{allcolors=[rgb]{0.0 0.0 0.75},linkcolor=[rgb]{0.75 0.05 0.05}}
\usepackage[dvipsnames]{xcolor}
\definecolor{darkblue}{RGB}{0,0,180}
\usepackage{mathtools}
\usepackage{amsmath}
\usepackage{amsfonts}
\usepackage{amssymb}
\usepackage{bbold}
\usepackage{bm}
\usepackage{orcidlink}
\usepackage{microtype}

\newcommand{\edit}[1]{{\color{darkblue}#1}}

\newcommand{\ket}[1]{|#1\rangle}

\newcommand{\bQ}{{\bf Q}}
\newcommand{\bD}{{\bf D}}

\newcommand{\bH}{{\bf H}}
\newcommand{\bB}{{\bf B}}
\newcommand{\bP}{{\bf P}}
\newcommand{\bR}{{\bf R}}
\newcommand{\bS}{{\bf S}}
\newcommand{\bM}{{\bf M}}
\newcommand{\bL}{{\bf L}}
\newcommand{\bJ}{{\bf J}}

\newcommand{\bh}{{\bf h}}

\newcommand{\sJ}{{\sf J}}
\newcommand{\sM}{{\sf M}}
\newcommand{\sD}{{\sf D}}

\newcommand{\sU}{{\sf U}}
\newcommand{\sH}{{\sf H}}
\newcommand{\sP}{{\sf P}}
\newcommand{\sR}{{\sf R}}

\newcommand{\sL}{{\sf L}}

\newcommand{\GF}{G_{\rm F}}
\newcommand{\exclude}[1]{{}}
\long\def\exclude#1{}

\begin{document}

\title{Slow and fast collective neutrino oscillations: Invariants and reciprocity}

\author{Damiano F.\ G.\ Fiorillo \orcidlink{0000-0003-4927-9850}} 
\affiliation{Niels Bohr International Academy, Niels Bohr Institute,
University of Copenhagen,\\ 2100 Copenhagen, Denmark}

\author{Georg G.\ Raffelt
\orcidlink{0000-0002-0199-9560}}
\affiliation{Max-Planck-Institut f\"ur Physik (Werner-Heisenberg-Institut), F\"ohringer Ring 6, 80805 M\"unchen, Germany}

\date{23 January 2023, \edit{Post-publication updates in blue, 15 January 2026}}

\begin{abstract}

The flavor evolution of a neutrino gas can show ``slow'' or ``fast'' collective motion. In terms of the usual Bloch vectors to describe the mean-field density matrices of a homogeneous neutrino gas, the slow two-flavor equations of motion (EOMs) are \hbox{$\dot\bP_\omega=(\omega\bB+\mu\bP)\times\bP_\omega$}, where $\omega=\Delta m^2/2E$, $\mu=\sqrt2\GF (n_\nu+n_{\bar\nu})$, $\bB$ is a unit vector in the mass direction in flavor space, and $\bP=\int d\omega\,\bP_\omega$. For an axisymmetric angle distribution, the fast EOMs are \smash{$\dot\bD_v=\mu(\bD_0-v\bD_1)\times\bD_v$}, where $\bD_v$ is the Bloch vector for lepton number, $v=\cos\theta$ is the velocity along the symmetry axis, $\bD_0=\int dv\,\bD_v$, and $\bD_1=\int dv\,v\bD_v$. We discuss similarities and differences between these generic cases. Both systems can have pendulum-like instabilities (soliton solutions), both have similar Gaudin invariants, and both are integrable in the classical and quantum case. Describing fast oscillations in a frame comoving with $\bD_1$ (which itself may execute pendulum-like motions) leads to transformed EOMs that are equivalent to an abstract slow system. These conclusions carry over to three flavors.

\end{abstract}

\maketitle

\section{Introduction}

The refractive energy shift of a neutrino in a medium of neutrinos or antineutrinos of its own flavor is twice that for a different flavor. In a seminal paper thirty years ago, Pantaleone showed that therefore neutrino-neutrino refraction is a many-body phenomenon \cite{Pantaleone:1992eq}. Shortly thereafter, in another foundational paper, Samuel found that neutrino flavor evolution feeding back on itself spawns intriguing forms of collective flavor evolution \cite{Samuel:1993uw}. While the community was somewhat slow at fully catching on to these insights, the idea is now standard that neutrino-neutrino refraction can strongly affect flavor-dependent neutrino transport, notably in supernova cores or other (neutrino-)dense environments. 

Some forty research papers on this topic have appeared in 2022 alone, making it impossible to do justice to this swelling body of literature in our brief introduction. So we merely refer to some early \cite{Duan:2009cd, Duan:2010bg} and more recent reviews \cite{Mirizzi:2015eza, Chakraborty:2016yeg,Tamborra:2020cul,Patwardhan:2023} that provide a glimpse of the richness of questions addressed in this quest. Much recent work is directed toward understanding space-time dependent solutions in the context of realistic astrophysical environments, including effects of collisions, or understanding possible limitations of the mean-field description.

We assume that the reader is largely familiar with this general subject and here return to far more elementary questions about the basic structure of the nonlinear flavor equations of motion (EOMs). In particular, we will juxtapose the basic reference case of ``slow'' (energy-dependent) flavor oscillations with the  ``fast'' (angle-dependent) case and discuss a number of striking similarities and a certain reciprocity between them. 

One key element of collective flavor evolution is the possible appearance of instabilities in the linearized equations, signifying soliton solutions (pendulum-like behavior) in the nonlinear regime. These were identified a long time ago in the simplest collective-oscillation case \cite{Samuel:1995ri, Duan:2005cp, Hannestad:2006nj, Duan:2007mv, Raffelt:2011yb}, which is a homogeneous and isotropic (or single-angle) neutrino gas with a nontrivial energy distribution. A similar behavior was recently identified in the simplest system of fast flavor conversion \cite{Johns:2019izj, Padilla-Gay:2021haz} that consists of a homogeneous and axisymmetric neutrino gas with a nontrivial angle distribution of the lepton-number flux. We here juxtapose these generic cases more systematically, discuss similarities and differences, and show a certain reciprocity between them.

To define the exact cases of comparison, we use the usual Bloch vectors to represent two-flavor density matrices. The simplest collective EOMs are
\begin{equation}\label{eq:EOM-slow}
    \dot\bP_\omega=(\omega\bB + \mu\bP)\times\bP_\omega,
\end{equation}
where $\omega=\Delta m^2/2E$ is the vacuum oscillation frequency, $\mu=\sqrt{2}\GF (n_\nu+n_{\bar\nu})$ is a measure of the neutrino-neutrino interaction strength, and $\bP=\int d\omega\,\bP_\omega$. Antineutrinos are included with negative $\omega$ and we use the flavor isospin convention: spin up means $\nu_e$ or $\bar\nu_\mu$, spin down means $\nu_\mu$ or~$\bar\nu_e$. We usually assume $|\omega|\ll\mu$, a condition that defines a ``neutrino-dense'' environment. Because flavor conversion is here driven by $\omega$, this case has come to be called ``slow flavor oscillations.''

If we interpret the Bloch vectors as classical angular momenta, these EOMs follow from the Hamiltonian\footnote{The evolution of any function $F$ on phase space is given by $\dot F=\{F,H\}$ where $\{{\cdot}{\,,\,}{\cdot}\}$ is a Poisson bracket. For classical angular momenta, $\{P^x_\omega,P^y_\omega\}=P^z_\omega$ and cyclic permutations, implying for example $\{\bP_\omega,\bB\cdot\bP_\omega\}=\bB\times\bP_\omega$.}
\begin{equation}\label{eq:Hamiltonian-slow}
   H_{\rm slow}=\bB\cdot\bP_1 + \frac{\mu}{2} \bP_0^2,
\end{equation}
where $\bP_0=\bP=\int d\omega\,\bP_\omega$ and $\bP_1=\int d\omega\,\omega\bP_\omega$.

The second case of comparison uses Bloch vectors for the density matrices of neutrinos minus that for antineutrinos, i.e., for the angle-dependent lepton-number occupation. Here spin up means $\nu_e-\nu_\mu$ occupation, spin down $\bar\nu_\mu-\bar\nu_e$ occupation. Assuming axial symmetry and using the fast flavor limit of vanishing neutrino mass splitting ($\omega=0$), the EOMs are
\begin{equation}\label{eq:EOM-fast}
    \dot\bD_v=\mu(\bD_0 - v\bD_1)\times\bD_v,
\end{equation}
where $v=\cos\theta$ is the mode velocity along the symmetry axis. In analogy to the slow case, we use the moments\footnote{This notation, that was also used in Appendix~A of Ref.~\cite{Pehlivan:2011hp}, is slightly ambiguous in that e.g.\ $\bD_0$ means the zeroth moment, not $\bD_v$ at $v=0$, but usually this should not lead to confusion.}
\begin{equation}
    \bD_n=\int_{-1}^{+1}dv\,v^n\,\bD_v.
\end{equation}
The fast EOMs derive from the Hamiltonian
\begin{equation}\label{eq:Hamiltonian-fast}
   H_{\rm fast}=\frac{\mu}{2}\left(\bD_0^2-\bD_1^2\right),
\end{equation}
once more using the canonical angular-momentum Poisson brackets for the $\bD_v$. Our aim is to compare the properties of these minimal slow and fast cases.

The slow Hamiltonian, or its quantum equivalent, arises in many other ``spin problems'' and in particular in the context of the BCS theory of superconductivity. Using Anderson's pseudo-spin formalism \cite{Anderson:1958zza}, a Cooper pair means ``spin down'' whereas unpaired electrons correspond to ``spin up,'' implying a perfect analogy. For example, pendulum-like behavior was discovered in this system, collective oscillation between the paired and unpaired state, in a nonequilibrium situation when dissipation was small \cite{Barkov:2004a, Barkov:2004b}. Many features of what we call collective neutrino oscillations were discussed in a long series of papers by Yuzbashyan and collaborators \cite{yuzbashyan2005solution, Yuzbashyan:2005-PRB, yuzbashyan2005prb, yuzbashyan:2006, yuzbashyan2008normal, Yuzbashyan:2018gbu}, where often the classical and quantum cases are clearly juxtaposed. We consider mostly the classical case (the mean-field description of neutrino oscillations) and the EOMs are central to our arguments rather than the Hamiltonians themselves. Indeed, we argue the reciprocity of the slow and fast system on the level of the EOMs, not on the basis of canonical transformations between the variables.

Returning to our discussion, as a final simplification we absorb the scale $\mu$ in the definition of dimensionless time and use dimensionless oscillation frequencies $w=\omega/\mu$.
Then finally we consider the two sets of EOMs
\begin{eqnarray}\label{eq:EOM-slow-A}
    \dot\bP_w&=&(w\bB\,   +\, \bP_0)\times\bP_w,
    \\[1ex]\label{eq:EOM-fast-A}
    \dot\bD_v&=&(\bD_0-v\bD_1)\times\bD_v,
\end{eqnarray}
where $-\infty<w<+\infty$ and $-1\leq v\leq+1$. We will continue to denote the Bloch vectors of the slow system as $\bP_w$ and those of the fast system as $\bD_v$. These equations look vaguely similar and it is the main point of our paper to show in which sense the two system are actually reciprocal to each other.

We begin our discussion in Sec.~\ref{sec:Fast-Slow} with the mapping between the fast and an abstract slow system and we identify and interpret the constraint for the slow system required to enable the reverse mapping. In Sec.~\ref{sec:Invariants} we develop the machinery of Lax vectors to identify the invariants and prove the integrability of both systems in the mechanical sense. Moreover, we discuss the equivalence to a system with only few effective degrees of freedom based on the roots of the spectral polynomial. In Sec.~\ref{sec:Normal-Modes} we juxtapose the dispersion relations for the fast and slow cases and show the consistency of the original mapping on the level of the linearized system. We also explain the equivalence of the normal modes to the roots of the $z$-component of the Lax vectors if the neutrinos begin in flavor eigenstates. Next, in Sec.~\ref{sec:Three-Flavor}, we extend the discussion to three flavors and derive the corresponding Gaudin invariants. We end with some general conclusions in Sec.~\ref{sec:Conclusion}. We address a number of additional topics in appendices. In App.~\ref{app:transformation} we explain the general transformation between moving frames in flavor space. In App.~\ref{sec:thermodynamic_limit} we discuss the dispersion relation in the continuum (thermodynamic) limit following earlier work on the Vlasov equation. Finally, in App.~\ref{sec:app_diagonalization}, we prove the integrability of the quantum version of the
fast system.

\section{Fast--Slow Reciprocity}
\label{sec:Fast-Slow}

\subsection{Transforming a General Fast System to a Constrained Slow System}

To compare  Eqs.~\eqref{eq:EOM-slow-A} and \eqref{eq:EOM-fast-A} we first identify a few conserved quantities. Integrating Eq.~\eqref{eq:EOM-fast-A} over $dv$ leads to $\dot\bD_0=0$ and thus reveals that $\bD_0$ is a fixed vector. Applying $\int v\,dv$ on both sides provides
\begin{equation}\label{eq:EOM-D1}
\dot\bD_1=(\bD_0+\bD_2)\times\bD_1.    
\end{equation}
The evolution of $\bD_1$ is an instantaneous precession and so its length is conserved. For those conditions where the motion is analogous to a gyroscopic pendulum, $\bD_0$ plays the role of gravity, whereas $\bD_1$ that of the moving radius vector that executes precessions and nutations. 

There is an evident reciprocity between $\bD_1$ moving relative to the fixed $\bD_0$ or conversely $\bD_0$ moving in a frame comoving with $\bD_1$. To express the fast oscillations of Eq.~\eqref{eq:EOM-fast-A} in this comoving frame we observe that Eqs.~\eqref{eq:EOM-fast-A} and \eqref{eq:EOM-D1} both engender differential rotations that can be added. Therefore, we can simply subtract the precession of the frame comoving with $\bD_1$ from the original precession of the individual $\bD_v$. Denoting Bloch vectors in the moving frame with a tilde, we thus find
\begin{equation}\label{eq:EOM-Dtilde}
    \partial_t\tilde\bD_v=-\bigl(v\tilde\bD_1+\tilde\bD_2\bigr)\times\tilde\bD_v.
\end{equation}
We present a more formal derivation of this transformation in Appendix~\ref{app:transformation}. Applying $\int v\,dv$ on both sides confirms that \smash{$\partial_t\tilde\bD_1=0$}. Moreover, applying  $\int dv$ on both sides reveals that \smash{$\partial_t\tilde\bD_0=-\tilde\bD_2\times\tilde\bD_0$} and so it is indeed $\tilde\bD_0$ that now is a moving object with conserved length. All vectors coincide between the two frames at $t=0$, so the spectra, stability conditions, and so forth are the same in both frames.

In the fast EOMs~\eqref{eq:EOM-fast-A}, the common precession around the fixed vector $\bD_0$ does not affect the internal motion of the system. One usually assumes that initially all Bloch vectors are oriented in the flavor direction, apart from a small seed that can trigger possible instabilities, i.e., a small deviation of collinearity among the $\bD_v$. We still assume that the conserved $\bD_0$ defines the $z$-direction. The flavor content of every mode is encoded in $D_v^z$ which does not depend on the common precession. Therefore, the questions of physical relevance are the same in the co-rotating system so that one can simply remove the first term of Eq.~\eqref{eq:EOM-fast-A} and finds the same physical answers. In this case, $\bD_0$ disappears from both Eqs.~\eqref{eq:EOM-fast-A} and \eqref{eq:EOM-D1} and thus never appears in the transformed Eq.~\eqref{eq:EOM-Dtilde}. Of course, it is physically clear that the internal motion of the $\bD_v$ relative to $\bD_1$ does not depend on the common precession of all $\bD_v$ and of $\bD_1=\int dv\,v\,\bD_v$ around $\bD_0$.

So far we have treated $w$ and $v$ as continuous parameters. However, sometimes it is more practical to think of discrete sets of Bloch vectors to avoid worrying about integration measures. In this spirit integrals $\int dv$ or $\int dw$ correspond to $\sum_v$ or $\sum_w$. With this understanding, the transformed EOMs~\eqref{eq:EOM-Dtilde} are identical in form to the slow EOMs~\eqref{eq:EOM-slow-A} with the identifications
\begin{equation}\label{eq:identification}
    \bB=-\frac{\tilde\bD_1}{D_1}, 
    \quad
    w=v D_1,
    \quad
    \bP_w=-v^2\tilde\bD_v\big|_{v=w/D_1},
\end{equation}
where \smash{$D_1=|\bD_1|=|\tilde\bD_1|$}, $-D_1\leq w\leq+D_1$, and
\begin{equation}\label{eq:identification-2}
    \bP_0=\sum_w \bP_w=-\sum_v v^2\tilde\bD_v=-\tilde\bD_2.
\end{equation}
These identifications are possible because in the new frame, the first moment \smash{$\tilde\bD_1$} is constant and acts as a fixed direction around which all polarization vectors precess with different frequencies.

\subsection{Constrained Slow System}

The slow system has a vector of conserved length that can play the role of a pendulum vector,\footnote{The integral in the continuous case would be interpreted in the principal value sense, provided $\bP_w$ is integrable at $w=0$. Notice, however, that $w$ corresponds to infinite $E$ and in a realistic medium would be exponentially suppressed. Moreover, Eq.~\eqref{eq:identification} contains a factor $w^2$ times a slowly-varying function.}
\begin{gather}\label{eq:Q}
    \bR=\bB-\sum_w \frac{\bP_w}{w}.
\end{gather}
Notice that \hbox{$\dot\bR=-\sum_w\frac{1}{w}\dot\bP_w=-\sum_w \frac{1}{w}(w\bB+\bP_0)\times\bP_w$} $=-\bB\times\bP_0-\bP_0\times(\sum_w\frac{1}{w}\bP_w-\bB+\bB)=\bP_0\times\bR$ and so indeed $\bR$ follows an instantaneous precession. Actually the Bloch vector $\bR$ thus defined is a special case of a Lax vector, a topic that we will discuss later in Sec.~\ref{sec:Lax-Vectors}.

However, the transformed variables of Eq.~\eqref{eq:identification} reveal \smash{$\bP_{-1}\equiv\sum_w \bP_w/w=-\sum_v(v\tilde\bD_v/D_1)=-\tilde\bD_1/D_1=\bB$} so that $\bR=0$. This case of a slow flavor pendulum has not been discussed in the literature, but of course Eq.~\eqref{eq:Q} always allows for the special case $\bR=0$. Notice that, if $\bR=0$ at the initial instant, it remains zero throughout the evolution. In our case, after transforming the fast EOMs, we have already shown that it is \smash{$\tilde\bD_0$} that is the moving quantity of conserved length. 

In the slow case with $\bR=0$, whether or not it derives from the fast--slow transformation, we should thus consider $\bP_{-2}\equiv\sum_w w^{-2}\bP_w$ instead of $\bR$. The derivative is $\dot\bP_{-2}=\sum_w w^{-2}\dot\bP_w=\sum_w w^{-2}(w\bB+\bP_0)\times\bP_w=\bB\times\bP_{-1}+\bP_0\times\bP_{-2}$. Because by assumption $\bP_{-1}=\bB$, the first term drops out and so indeed $\dot\bP_{-2}=\bP_0\times\bP_{-2}$ is a precession equation that conserves the length of $\bP_{-2}$.

We can also reverse the transformation and start from the slow EOMs~\eqref{eq:EOM-slow-A} and go to a frame that moves with the vector that stays of constant length. This is either $\bR$ or $\bP_{-2}$, but in both cases the evolution is driven by $\bP_0$. Following the same logic as in the earlier transformation, the moving EOMs are $\partial_t\tilde\bP_w=w\tilde\bB\times\tilde\bP_w$. Next we can use Eq.~\eqref{eq:Q} to eliminate $\tilde\bB$ and find with the identification $v=w$ and \smash{$\bD_v=-\tilde\bP_w/w^2$} the equivalent EOMs
\begin{equation}\label{eq:slow-to-fast}
    \dot\bD_v=-v(\bD_1-\bR)\times\bD_v.
\end{equation}
This has the form of a fast system, but only if $\bR=0$ of the original slow system.

Therefore, in both directions of the transformation one finds that the fast EOMs are equivalent to a specific slow system where $\bR=0$.

The meaning of this constraint becomes more obvious for a system of $N$ discrete modes $\bD_v$ in the fast system. After the transformation we obtain $N$ slow modes $\bP_w$ and some combination of $\bD_v$ playing the role of the fixed vector $\bB$. The additional degree of freedom is absorbed by the constraint $\bR=0$ so that the slow system also has $N$ Bloch-vector degrees of freedom.

Therefore, we find that a slow system can be mapped to the same form as a fast system if the vector $\bR=0$. However, more generally, we can transform the EOMs to a frame co-rotating around the $z$ axis with frequency $w_{\rm c}$, where the EOMs are
\begin{equation}
\dot{\bP}_w=\left[(w-w_{\rm c})\bB+\bP_0\right]\times\bP_w
\end{equation}
and perform the same steps as before. In this frame, we need to require that 
\begin{equation}
\bR_{\rm c}=\bB-\sum_w\frac{\bP_w}{w-w_{\rm c}}=0.
\end{equation}
This expression has the same form as a Lax vector Eq.~\eqref{eq:Lax-slow} of this system. In other words, the slow-to-fast mapping can be done if there exists any vanishing Lax vector $\bL_u$ with real $u$.

\subsection{Conserved Energy}

The slow and fast EOMs derive respectively from the classical Hamiltonians Eqs.~\eqref{eq:Hamiltonian-slow} and~\eqref{eq:Hamiltonian-fast} if we interpret the Bloch vectors as classical angular momenta. The two terms in $H_{\rm fast}$ are separately conserved and indeed, the same motion obtains if we leave out the first piece $\frac{1}{2}\bD_0^2$, except for an overall precession around $\bD_0$. In the slow case, on the other hand, the two pieces exchange energy in the course of the motion. For the pendulum solutions, the first term plays the role of potential energy, the second term the role of kinetic energy.

After mapping the fast system on a slow one, we recall that $\bB\sim\bD_1$, $\bP_1\sim\bD_3$, and $\bP_0\sim\bD_2$. Therefore one expects that the quantity
\begin{equation}\label{eq:shifted_frame_hamiltonian}
    \bD_1\cdot\bD_3+\frac{1}{2}\bD_2^2
\end{equation}
is conserved as one can verify using the fast EOMs~\eqref{eq:EOM-fast-A}. This is one of many invariants of the fast system.

\section{Invariants and Integrability}

\label{sec:Invariants}

\subsection{Lax Vectors}
\label{sec:Lax-Vectors}


The slow EOMs~\eqref{eq:EOM-slow-A} admit an infinite set of constants of the motion which are best expressed in terms of the so-called Lax vectors \cite{yuzbashyan2005prb,yuzbashyan2008normal}
\begin{equation}\label{eq:Lax-slow}
    \bL_u=\mathbf{B}+\sum_{w\not=u}\frac{\mathbf{P}_{w}}{u-w},
\end{equation}
where in general $u$ is a complex parameter, whereas $w$ is understood as a set of discrete frequencies. The time derivative is found by inserting Eq.~\eqref{eq:EOM-slow-A} for $\dot\bP_w$ on the right-hand side, leading to
\edit{\begin{equation}
    \dot\bL_u=(u\bB+\bP_0)\times\bL_u+\bB\times \bP_u
\end{equation}
and where $\bP_0=\sum_w\bP_w$. For $u\neq w$, i.e., not coinciding with any of the frequencies of the original system, the last term disappears. In these cases, this is a formal precession equation with complex frequency $u$ and implies that the length of $\bL_u$ is conserved, i.e., $\partial_t \bL_u^2=0$, where $\bL_u^2$ is a complex number.

Instead, when $u$ coincides with one of the frequencies $w$, the last term is present, and the motion of $\bL_u$ is not a pure precession. However, one can easily verify that in both cases, $\bL_u\cdot\bP_u$ is conserved.

The somewhat discontinuous behavior of the Lax vectors, which perform a pure precession only for $u\neq w$, can be cured by introducing a modified definition
\begin{equation}\nonumber
    \tilde{\bL}_u=\prod_w(u-w)\left[\bB+\sum_w\frac{\bP_w}{u-w}\right].
\end{equation}
These modified Lax vectors continuously obey
\begin{equation}\nonumber
    \dot{\tilde{\bL}}_u=(u\bB+\bP_0)\times \tilde{\bL}_u.
\end{equation}
On the other hand, for $u$ equal to one of the frequencies of the original system, the modified Lax vectors coincide with the original polarization vector $\tilde{\bL}_w=\bP_w$, so that the invariants $\tilde{\bL}_w\cdot\bP_w$ are not independent constants of motion. (Still, for all other continuous values of $u\neq w$, \smash{$|\tilde{\bL}_u|^2$} is a non-trivial invariant.)
}

For the fast system, we find that the corresponding Lax vectors are
\begin{equation}\label{eq:Lax-fast}
    \bL_u=\sum_{v\not=u}\frac{v\bD_{v}}{u-v}.
\end{equation}
In analogy, the time derivative is found by inserting Eq.~\eqref{eq:EOM-fast-A} for $\dot\bD_v$ on the right-hand side, leading to
\edit{\begin{equation}\label{eq:Lax-fast-EOM}
    \dot\bL_u=(\bD_0-u\bD_1)\times\bL_u-u\bD_1\times\bD_u
\end{equation}}%
with $\bD_0=\sum_v\bD_v$ and $\bD_1=\sum_vv\bD_v$. So here as well, the complex length of $\bL_u$ is conserved for any complex~$u$, \edit{except for $u$ equal to one of the velocities of the original system. In that case, the non-trivial invariants are the products $\bL_v\cdot \bP_v$, the so-called Gaudin invariants, which we now introduce in more detail.}

\subsection{Gaudin Invariants}
\label{sec:Gaudin-Invariants}

\subsubsection{Slow System}

\exclude{
Defining instead the Lax vectors for the same discrete set $w$ of frequencies we may define the variables
\begin{equation}
    \bQ_w=\bB+\sum_{w'\not=w}\frac{\bP_{w'}}{w-w'},
\end{equation}
where now every $\bQ_w$ obeys the same EOM as $\bP_w$, implying that $\bQ_w\cdot\bP_w$ is invariant. 
}

The Lax vectors provide a continuous infinity of invariants of the motion, which however are not linearly independent. However, considering the Lax vectors only for the same discrete set $w$ of frequencies, then there is a real Lax vector $\bL_w$ for every $\bP_w$.
It provides the Gaudin invariant or Gaudin Hamiltonian
\cite{Gaudin:1976sv}
\begin{equation}\label{eq:Gaudin-Hamiltonian}
    I_w=\bL_w\cdot\bP_w=\bB\cdot\bP_w+\sum_{w'\not=w}\frac{\bP_{w}\cdot\bP_{w'}}{w-w'}.
\end{equation}
This is a conserved quantity of our original system, but can also be interpreted as a Hamiltonian that defines a new system of a ``central'' spin $\bP_w$ coupled to an external $\bB$ field and to a set of ``environmental'' spins, the ``central spin problem'' (see e.g.\ Ref.~\cite{yuzbashyan2005solution}).

The Gaudin invariants generate all the other integrals of motion. One example is 
\begin{equation}\label{eq:C0}
    \sum_w I_w=\bB\cdot\sum_w\bP_w=\bB\cdot\bP_0
\end{equation}
as can be seen immediately from the antisymmetry of the second term in Eq.~\eqref{eq:Gaudin-Hamiltonian}
under exchange $w\leftrightarrow w'$. Likewise, the Hamiltonian of the original slow system can be expressed as\footnote{This is seen if we consider
\begin{eqnarray}
&& \sum_{\substack{w,w'\\w\not=w'}}\frac{w\bP_w\cdot\bP_{w'}}{w-w'}=\sum_{\substack{w,w'\\w\not=w'}}\frac{(w-w')\bP_w\cdot\bP_{w'}+w'\bP_w\cdot\bP_{w'}}{w-w'}
    \nonumber\\
&&{}=\sum_{w,w'}\bP_w\cdot\bP_{w'}-\sum_w\bP_w^2+ \sum_{\substack{w,w'\\w\not=w'}}\frac{w'\bP_w\cdot\bP_{w'}}{w-w'}
\nonumber\\
&&{}=\bP_0\cdot\bP_0-\sum_w\bP_w^2-\sum_{\substack{w,w'\\w\not=w'}}\frac{w\bP_w\cdot\bP_{w'}}{w-w'},\nonumber
\end{eqnarray}
where we have interchanged the summation variables in the last expression, leading to the minus sign. Taking this term to the left-hand side finally reveals that the original left-hand side is simply $\frac{1}{2}\left(\bP_0\cdot\bP_0-\sum_w\bP_w^2\right)$.
}
\begin{equation}\label{eq:C1}
    H_{\rm slow}=\sum_w \bigl(wI_w+\frac{1}{2}\bP_w^2\bigr)=\bB\cdot\bP_1+\frac{1}{2}\bP_0^2.
\end{equation}
Notice that $\sum_w\bP_w^2$ is separately conserved and thus could be left out without changing the EOMs. This term is included to make up for the piece $w=w'$ that is missing in sums involving $1/(w-w')$.

A generalization of these invariants is given by the expressions\footnote{For additional expressions see Appendix~A of Ref.~\cite{Pehlivan:2011hp}. Their Eq.~(A3) correspond to our Eq.~\eqref{eq:Cn} with their $\mu\to1/2$.}
\begin{subequations}
    \begin{eqnarray}\label{eq:Cn}
    C_n&=&\sum_w\bigl( w^n I_w+\frac{1}{2}n w^{n-1}\bP_w^2\bigr)
    \\
    &=&\bB\cdot\bP_n+\frac{1}{2}\sum_{m=0}^{n-1}\bP_m\cdot\bP_{n-1-m},
\end{eqnarray}
\end{subequations}
where $C_0$ was already given in Eq.~\eqref{eq:C0} and $C_1$ is $H_{\rm slow}$ given in Eq.~\eqref{eq:C1}.

\subsubsection{Fast System}

To find the Gaudin invariants of the fast system, once more we consider the Lax vectors $\bL_v$
for the same set of discrete velocities $v$ as the original fast system.
Once more they follow the same EOMs as the original set $\bD_v$ and so we find the Gaudin invariants
\begin{equation}
    I_v=\bD_v\cdot\bL_v=\sum_{v'\not=v}\frac{v'\bD_{v'}\cdot\bD_v}{v-v'}.
\end{equation}
A trivial conserved quantity is
\begin{equation}
C_0=\sum_v I_v+\frac{1}{2}\sum_v \bD_v^2=-\frac{1}{2} \bD_0^2.
\end{equation}
The next invariant
\begin{equation}
C_1=\sum_v v I_v=0
\end{equation}
vanishes identically.

Further nontrivial conserved quantities are obtained for $n\geq2$
\begin{subequations}
    \begin{eqnarray}\label{eq:Cn-fast}
    C_n&=&\sum_v\Bigl[v^n I_v+ \frac{1}{2} v^n(n-1)\bD_v^2\Bigr]
    \\
    &=&\frac{1}{2}\sum_{m=1}^{n-1}\bD_m\cdot\bD_{n-m}.
\end{eqnarray}
\end{subequations}
In terms of this chain of invariants, the fast Hamiltonian Eq.~\eqref{eq:Hamiltonian-fast} is
\begin{equation}
H_{\mathrm{fast}}=-(C_0+C_2)=\frac{1}{2}\bD_0^2-\frac{1}{2}\bD_1^2.
\end{equation}
On the other hand, the Hamiltonian in the frame comoving with $\bD_1$ that was given in  Eq.~\eqref{eq:shifted_frame_hamiltonian} is the same as $C_4$. The invariants of the system can be written in many ways, each providing a different glance on its properties.

\subsection{How Many Invariants per Mode?}
\label{sec:HowMany}

We have identified two invariants for every mode, the length of the Bloch vector and the Gaudin invariant. The latter is the scalar product of the Bloch vector and the corresponding Lax vector. So for the fast system, we have $\bD_v^2$ and $\bD_v\cdot\bL_v$ as invariants. \edit{On the other hand, we have in addition a whole new continuous set of conserved invariants $|\bL_u|^2$ for $u$ not belonging to the set of original velocities.} While it is physically obvious that the system cannot have another independent invariant per mode, it is instructive to show this explicitly.

\edit{
Following the definition of the Lax vector in Eq.~\eqref{eq:Lax-fast}, its square is
\begin{equation}
   \mathbf{L}_u^2=\sum_{v_1,v_2}\mathbf{D}_{v_1}\cdot\mathbf{D}_{v_2}\,
    \frac{v_1 v_2}{(u-v_1)(u-v_2)}.
\end{equation}
We separate from this equation the terms with $v_1=v_2$ and those with $v_1\neq v_2$: for the latter, we modify the right-hand side with the identity
\begin{equation}
    \frac{v_1 v_2}{(u-v_1)(u-v_2)}=\frac{v_1 v_2}{v_1-v_2}\left(\frac{1}{u-v_1}-\frac{1}{u-v_2}\right)
\end{equation}
so that
\begin{eqnarray}\label{eq:non_independent_conservation}
    \mathbf{L}_u^2&=&\sum_{v_1}\frac{\bD_{v_1}^2 v_1^2}{(u-v_1)^2}+2\sum_{v_1,v_2}\frac{\mathbf{D}_{v_1}\cdot\mathbf{D}_{v_2}\,v_1 v_2}{(v_1-v_2)(u-v_1)}
    \nonumber\\
    &=&\sum_{v_1}\frac{D_{v_1}^2 v_1^2}{(u-v_1)^2}+2\sum_{v_1} I_{v_1}\frac{v_1}{u-v_1}.
\end{eqnarray}
Therefore, the set of invariants $\bL_u^2$ is indeed a linear combination of the Gaudin invariants $I_v$ and the original lengths of the Bloch vectors, without introducing any additional constant of motion.
}

\subsection{Integrability}

The existence of two constants of motion for each of the Bloch vectors makes the system completely integrable in the mechanical sense. First of all, the nature of the motion as a precession (or in matrix form, the commutator structure) reveals that the length of each Bloch vector is conserved, reducing the cartesian variables to two polar coordinates. Therefore, the motion involves periodic variables. In addition, there is another conserved quantity for each $\bP_w$ given, for example, by the Gaudin invariants. (The nature of the fast case as a special case of the slow one makes redundant a separate discussion of the former.)

Therefore, the system has only as many underlying variables as independent Bloch vectors, i.e., the number of different discrete $w$. Indeed, one can introduce canonical variables which separate the Hamilton-Jacobi equations as those frequencies $\omega^*$ for which the vectors $\mathbf{L}_{\omega^*}$ are along the $z$-axis, and their canonical conjugate variables, namely the $z$ component of $\mathbf{L}_{\omega^*}$. An explicit solution for any initial condition can then be constructed following, e.g., Refs.~\cite{yuzbashyan2005solution, kuznetsov1992quadrics}. 

When the system is quantized, the integrability of the equations shows up in the possibility of constructing all of the eigenstates of the Hamiltonian analytically by the Bethe ansatz. 
(See Appendix~\ref{sec:app_diagonalization} for the integrability of the quantum system.)

\subsection{Degenerate Solutions}

The large number of conserved quantities implies that the motion in the $2N_{\rm tot}$-dimensional phase space -- $N_{\rm tot}$ being the number of Bloch vectors -- is restricted to an $N_{\rm tot}$-dimensional torus. However, depending on the set of frequencies $\omega_i$ and initial $\bP_{i}$, the motion of the system can be constrained to a reduced phase space instead of ergodically filling all of the torus. Such solutions were called ``degenerate'' in Ref.~\cite{yuzbashyan2005solution}, whereas in a previous study by one of us \cite{Raffelt:2011yb} they were dubbed N-mode coherent solutions. Either way, the solutions $\bP_i(t)$ with $i=1,\ldots,N_{\rm tot}$ would be linear combinations of a smaller set of functions $\bJ_i(t)$ with $i=1,\ldots,N<N_{\rm tot}$ and in this sense the full set $\bP_i(t)$ moves coherently rather than each of them independently. The $\bJ_i(t)$ were termed ``auxiliary spins'' in Ref.~\cite{yuzbashyan2005solution} or ``carrier modes'' in Ref.~\cite{Raffelt:2011yb}.

To show that this phenomenon can indeed exist, we first reverse the problem and introduce a fictitious system of a smaller number $N<N_{\rm tot}$ of Bloch vectors $\bJ_i$ with frequencies $\chi_i$ obeying an analogous slow EOM with the same unit vector~$\bB$
\begin{equation}
    \dot\bJ_i=(\chi_i \bB+\mu \bJ)\times\bJ_i,
\end{equation}
where $\bJ=\sum_{i=1}^{N}\bJ_i$. (We here return to dimensionful frequencies and coupling strength $\mu$.) This is yet another slow system and as such integrable.

Therefore, we can construct Lax vectors on an arbitrary set of frequencies $\omega_i$ with $i=1,\dots,N_{\rm tot}>N$ and suppose that none of the chosen $\omega_i$ coincides with any of the $\chi_i$. This set of Lax vectors of the fictitious system is
\begin{equation}
    \bL_i=\bB+\mu\sum_{j=1}^N \frac{\bJ_j}{\omega_i-\chi_j}
\end{equation}
and as such obeys the EOMs
\begin{equation}
    \dot\bL_{i}=(\omega_i\bB+\mu\bJ)\times\bL_i.
\end{equation}
Let us assume that the \{$\bL_i$, $\omega_i$\} are fixed such that the matching condition $\bL=\bJ$ is obeyed. This can be achieved by another factor depending on $i$ that adjusts the arbitrary length of $\bL_i$ and/or the choice of $\omega_i$. The new system $\bL_i$ is a higher-dimensional slow system with solutions $\bL_i(t)$ that depend on $\bB$ and $N<N_{\rm tot}$ independent functions $\bJ_i(t)$ with $i=1,\ldots,N$.

More difficult is the reverse problem where we begin with a given system \{$\bP_i,\omega_i$\} with $i=1,\ldots,N_{\rm tot}$ and ask for its true dimensionality. One numerical diagnostic requires to solve the EOMs for some chosen time interval and determine the components of the Gram matrix, \smash{$G_{ij}=\int_{t_1}^{t_2}\bP_i(t)\cdot\bP_j(t)$} \cite{Raffelt:2011yb}. Its rank (the number of independent eigenvalues) reveals the number of linearly independent underlying functions.

However, this dimensionality can also be directly determined from the initial conditions without solving the EOMs \cite{yuzbashyan2005solution}. The key information is encoded in the complex length of the Lax vector $\bL_u^2$. If it vanishes for a real $u$ at the initial time, it must vanish at all times, which means that one of the polarization vectors can be expressed in terms of the others, and therefore the number of degrees of freedom drops by~one. 

Indeed, the root diagram of $\bL_u^2$, the generally complex solutions $u_i$ of $\bL_u^2=0$, provides a lot of crucial information about a given system~\cite{yuzbashyan2005solution}.
Formally one defines the ``spectral polynomial'' of our system by
\begin{equation}
    Q(u)=\bL_u^2\,p_u^2
    \quad\hbox{where}\quad
    p_u=\prod_{i=1}^{N_{\rm tot}}(u-\omega_i).
\end{equation}
Here, $p_u$ is the product of the denominators of the Lax vectors and so, $Q_u$ is the numerator of $\bL_u^2$ and as such a polynomial in $u$ of order $2N_{\rm tot}$. For all real $u$, we have $Q_u\geq0$, which implies that the $2N_{\rm tot}$ zeroes are pairs of complex conjugate solutions. As a consequence, real zeros come in pairs of degenerate solutions.

In the fast flavor case, the constant vector $\bB$ is absent from the Lax vector Eq.~\eqref{eq:Lax-fast}, but on the other hand there is a factor $u$ in the numerator, so the spectral polynomial once more has degree $2N_{\rm tot}$ if there are $N_{\rm tot}$ Bloch vectors $\bD_v$. However, one trivial real double root is now $u=0$, so there are at most $N_{\rm tot}-1$ independent degrees of freedom, whereas one is eliminated, corresponding to the conserved~$\bD_0$.

In summary, the number $N$ of pairs of complex conjugate roots of the spectral polynomial is the number of independent underlying degrees of freedom. For an arbitrary set $\{\bP_i,\omega_i\}$, very special initial conditions are required to obtain a reduced number of degrees of freedom. In the continuum of possible initial conditions, only a discrete set of measure zero fulfills this condition~\cite{yuzbashyan2005solution}. However, in the neutrino context we consider very special initial conditions, i.e., all Bloch vectors are initially collinear with the flavor direction. In this case, the situation is reversed in that there are very few independent degrees of freedom, providing soliton solutions. 

\section{Normal Modes and Lax Vectors}

\label{sec:Normal-Modes}

In the context of dense neutrino environments, one usually imagines that they were produced in flavor eigenstates and that the in-medium mixing angles are small. So for both slow and and fast oscillations, one typically studies initial conditions where all Bloch vectors are collinear with the flavor direction that we identify with the $z$-direction in flavor space. Analogous assumptions were made in the cited discussions of the BCS Hamiltonian where the initial state consists of paired or unpaired electrons, but not initially of coherent superpositions of paired with unpaired states \cite{yuzbashyan2005solution, Yuzbashyan:2005-PRB, yuzbashyan2005prb, yuzbashyan:2006, yuzbashyan2008normal, Yuzbashyan:2018gbu}. 

Any evolution away from this initial state begins with small excursions from the $z$-directions, lending itself to a linearized analysis and to the identification of possible run-away modes that can lead to large flavor-conversion effects. Eventually these modes become nonlinear, representing what were called ``pendulum solutions'' in the neutrino literature or ``solitons'' in the BCS literature. We juxtapose the fast and slow cases in the context of a linear normal-mode analysis and we highlight the connection in both cases between the linear normal-mode analysis and the Lax-vector analysis.

\subsection{Normal-Mode Analysis}

\subsubsection{Slow System}

The fast and slow systems can both be analyzed in the linear regime of small off-diagonal components of the density matrices. This approach makes sense in the usual picture where all Bloch vectors start essentially collinear except for a small seed to trigger run-away modes. For both systems, the starting point are the ``spectra''
\begin{equation}
    G_v=D_v^z|_{t=0}
\quad\hbox{and}\quad
    g_w=P_w^z|_{t=0}
\end{equation}
that define the systems. This includes both continuous functions or collections of discrete modes that are represented by collections of $\delta$ functions at different values of $w$ or $v$. In this case, integrations over the spectra are effectively summations over discrete modes.

The most general space-time dependent linearization and concomitant dispersion relation was developed in Ref.~\cite{Izaguirre:2016gsx}, but here it is somewhat more transparent to linearize our simple systems directly. For the slow case we may follow the first study of linear stability analysis \cite{Banerjee:2011fj} and note that initially $P_w^{xy}\equiv P_w^{x}-i P_w^y$ is small, whereas $P_w^z=g_w$ is at its starting value.\footnote{From the relation to the density matrix $\varrho_w=\frac{1}{2}\bP_w\cdot{\bm\sigma}$ we observe that $P_w^{xy}$ is the ``upper right'' entry of $\varrho_w$. We could also use the lower-left off-diagonal element, the complex conjugate of $P_w^{xy}$, leading however to the same information.}
Then the linearized version of Eq.~\eqref{eq:EOM-slow-A} is 
\begin{equation}\label{eq:schrodinger}
(i\partial_t+P_0^z-w B^z)P_w^{xy}=-g_w\int dw' P_{w'}^{xy},
\end{equation}
where $P_0^z=\int dw\,g_w$. We have explicitly included $B^z$ for later comparison with $D_1^z$. While $\bB$ is defined as a unit vector collinear with the $z$-direction, both $B^z=\pm1$ are in principle possible.

A collective normal mode is of the form $P_w^{xy}\propto g_w Q_w e^{-i\Omega t}$, where $\Omega$ is the eigenfrequency that can be complex. If $\bP_w$ with length $|g_w|$ is expressed in polar coordinates $(\vartheta_w,\varphi_w)$, we have $Q_w=\sin\vartheta_w\,e^{-i\varphi_w}$. The linear EOMs imply the eigenfunction
\begin{equation}\label{eq:eigenvectors}
    Q_w=\frac{A}{w B^z-P_0^z-\Omega},
\end{equation}
where $A$ does not depend on $w$. Inserting this result back into the linear EOM reveals that the eigenfrequency is fixed by the condition
\begin{equation}\label{eq:slow-eigenvalue}
    \int dw\,\frac{g_w}{w B^z-P_0^z-\Omega}=1,
\end{equation}
Recall that while $\bP_0$ moves, its projection $P_0^z=\int dw\,g_w$ on $\bB$ is conserved. 

If the spectrum consists of $n$ discrete modes, one finds $n$ solutions that can be real or pairs of complex conjugate ones. If $g_w$ is continuous and nonzero for all $w$, except perhaps some crossings, the poles in the integral prevent any real solution of the ``dispersion relation'' Eq.~\eqref{eq:slow-eigenvalue}. The singularity associated with the vanishing of the denominator in Eq.~\eqref{eq:slow-eigenvalue} requires special care in the interpretation, although it leads to no practical difficulties for the special problem we have at hand, namely identifying the instabilities of the system. We discuss this issue in Appendix~\ref{sec:thermodynamic_limit}. For the discussion in the main text, we limit only to complex frequencies, for which there is no pole in the integration; this is enough to understand whether there are instabilities.



If $\Omega=\omega_{\rm P}+i\Gamma$ is complex, the condition of Eq.~\eqref{eq:slow-eigenvalue} consists of a real and imaginary part that are explicitly
\begin{subequations}
\begin{eqnarray}
    \int dw\,\frac{g_w(wB^z-P_0^z-\omega_{\rm P})}{(wB^z-P_0^z-\omega_{\rm P})^2+\Gamma^2}&=&1,
    \\[1.5ex]
    \int dw\,\frac{g_w\Gamma}{(w B^z-P_0^z-\omega_{\rm P})^2+\Gamma^2}&=&0.
\end{eqnarray}
\end{subequations}
The second equation implies that if there is any solution with $\Gamma\not=0$, then there are two unstable solutions $\pm \Gamma$, i.e., an exponentially growing and shrinking one, a point that is of course obvious in the discrete case anyway. Moreover, for $\Gamma\not=0$, the second equation can only be true if $g_w$ has regions of positive and negative sign, an argument that also applies to both the discrete and continuum case. Unstable solutions require a ``spectral crossing'' as a necessary condition, which is also sufficient in the more general case of inhomogeneous solutions \cite{Morinaga:2021vmc, Dasgupta:2021gfs}.

\subsubsection{Fast System}

For the fast case, we follow the same steps or use directly the explicit derivation in Ref.~\cite{Padilla-Gay:2021haz}. Here the linear EOMs~are
\begin{equation}\label{eq:schrodinger-fast}
(i\partial_t-D_0^z+v D^z_1)D_v^{xy}=v G_v\int dv' v' D_{v'}^{xy},
\end{equation}
with the additional constraint $\sum_v D_v^{xy}=0$ because the $z$-direction is defined to be the direction along $\bD_0$. We will return to this question below. The eigenfunction is
\begin{equation}\label{eq:eigenvectors-fast}
    Q_v=\frac{A\,v}{v D_1^z-D_0^z+\Omega},
\end{equation}
providing the condition
\begin{equation}\label{eq:fast-eigenvalue}
    \int_{-1}^{+1}dv\,\frac{v^2 G_v}{v D_1^z-D_0^z+\Omega}=1,
\end{equation}
where $D_0^z=\int dv G_v$ and  $D_1^z=\int dv v G_v$. Notice that $D_0^z$ only shifts the real part of $\Omega$ and thus has no impact on the stability and was left out in Ref.~\cite{Padilla-Gay:2021haz}. In this sense one can use a co-rotating frame where $\bD_0$ drops out from the original EOMs. However, to avoid jumping between too many moving frames, we here keep all terms explicitly.

To compare Eqs.~\eqref{eq:slow-eigenvalue} and \eqref{eq:fast-eigenvalue} we recall that $D_1=|\bD_1|=|D_1^z|$ is positive, whereas $D_1^z=\int dv\, G_v$ can be both positive or negative. If we absorb $D_0^z$ and $P_0^z$ in the real part of $\Omega$, Eq.~\eqref{eq:slow-eigenvalue} follows from Eq.~\eqref{eq:fast-eigenvalue} with the substitution $D_1^z\to -B^z D_1$, $v\to w/D_1$, $v^2 G(v)\to -g(w)$, meaning $g(w)=-(w/D_1)^2 G(w/D_1)/D_1$, and the limits of integration are $\pm D_1$, or alternatively one can say that $g(w)=0$ for $|w|>D_1$. These identifications correspond to Eqs.~\eqref{eq:identification} and \eqref{eq:identification-2}, but now for a continuous system that involves a factor $D_1$ as a Jacobian from the $v\to w$ transformation.

Equation~\eqref{eq:fast-eigenvalue} is trivially solved by the real solution $\Omega=D_0^z$. The corresponding eigenvector $Q_v$ from Eq.~\eqref{eq:eigenvectors-fast} is independent of $v$, and therefore seems to represent a uniform rotation of all polarization vectors, which however we had excluded in the beginning by the condition $\int dv D^{xy}_v=0$. However, the 
$\Omega=D_0^z$ is not spurious, but actually can be constructed explicitly. Notice that for $\Omega=D_0^z$, the eigenvector $Q_v$ can possess an arbitrary contribution from the Bloch vector with $v=0$, which automatically satisfies the linearized EOM~\eqref{eq:schrodinger-fast}. Therefore, the eigenvector for the mode $\Omega=D_0^z$ is
\begin{equation}
Q_v=\frac{1}{D_1^z}\left[1-\frac{D^z_0}{G_v}\,\delta(v)\right],
\end{equation}
which correctly satisfies the constraint $\int dv D^{xy}_v=0$. 

Physically, this mode corresponds to all Bloch vectors aligned along a direction slightly tilted from the $z$-axis, while the $v=0$ vector is tilted in the opposite way so that $\bD_0$ is still along the $z$-axis. Since $\bD_{v=0}$ does not contribute to $\bD_1$ (and in fact all $\bD_n$ with $n>0$), the latter is still aligned with all Bloch vectors with $v\neq0$, which therefore just precess around $\bD_0$, whereas the polarization vector with $v=0$ automatically satisfies the same pure precession motion since $\dot{\bD}_{v=0}=\bD_0\times\bD_{v=0}$. Therefore, the fast system always admits a stable, uniformly rotating mode, in which all the moments $\bD_n$ with $n>0$ are slightly tilted from the $z$ axis. The existence of this mode is crucially connected with the conservation of $\bD_0$.

By that same token, any function $g_w$ derived from $G_v$ as a fast-to-slow conversion fulfills
\begin{equation}\label{eq:constraint_linear}
    \int dw\,\frac{g_w}{w}=\pm1,
\end{equation}
where the sign depends on $B^z$. On the other hand, starting with a generic slow system with any function $g_w$, normally it will not fulfill this constraint which indeed corresponds exactly to the constraint derived in the previous section that the vector $\bR$ vanishes, which is a necessary condition for a slow system to be mapped to a fast one. We now understand the physical origin of this issue: a fast system always admits a conserved vector $\bD_0$ because of its rotational invariance, and in turn it admits a uniformly precessing mode. A slow system can only be mapped to a fast system if it accidentally also admits a conserved vector. In the slow system case, this conservation is not protected by any fundamental symmetry, and therefore it only arises if the condition $\bR=0$ is accidentally satisfied.

For later reference, we finish by providing an alternative form of the fast eigenvalue equation. To this end, we first absorb $D_0^z$ in the definition of $\Omega$ in Eq.~\eqref{eq:fast-eigenvalue}. On the right-hand side, we use the manipulation
\begin{eqnarray}
    1&=&\frac{\int dv\, v G_v}{D_1^z}=\frac{1}{D_z^z}\int dv\, G_v\frac{v(v D_1^z+\Omega)}{v D_1^z+\Omega}
    \nonumber\\[1ex]
    &=&\int dv\, G_v\frac{v^2+v\,\Omega/D_1^z}{v D_1^z+\Omega},
\end{eqnarray}
providing the alternative form
\begin{equation}\label{eq:fast-eigenvalue-linear}
    \int_{-1}^{+1}dv\,\frac{v\,G_v}{v D_1^z-D_0^z+\Omega}=0,
\end{equation}
where we have restored the original meaning of $\Omega$.

\subsection{Sign of Spectral Crossing}

A complex solution for $\Omega$ requires a spectral crossing as discussed earlier. In the slow case, it is well known that in addition the crossing must be positive, meaning that $g_w$ as a function of $w$ must cross from negative to positive values of $g_w$ \cite{Dasgupta:2009mg}. This condition applies for $\bB$ oriented in the positive $z$-direction or rather, for $\bB$ defining what is the positive $z$-direction. The reason for this condition is that the potential energy of the slow Hamiltonian in Eq.~\eqref{eq:Hamiltonian-slow} is already minimal for a negative crossing and cannot be lowered by the $\bP_w$ moving away from their initial position. Of course, a system with a positive crossing need not be unstable---it could be stuck in the ``sleeping top'' position.

Therefore, also in the fast system the crossing should be positive measured relative to the direction defined by $D_1^z=\int dv\,G_v$. For the explicit examples constructed in Ref.~\cite{Padilla-Gay:2021haz}, this is indeed the case. While the crossings of $G_w$ shown in their Fig.~1 look negative, also $D_1^z<0$. So a positive crossing in this sense is a necessary condition, but also in the fast case not a sufficient one, as seen in example~A of that paper.

\subsection{Normal Modes from Lax Vectors}

The eigenvalue equation for normal modes presented earlier can be derived in a much simpler way using the Lax vectors of a given system. We focus on the example of the fast system, but the proof proceeds identically for the slow system as well. The Lax vector defined in Eq.~\eqref{eq:Lax-fast} obeys Eq.~\eqref{eq:Lax-fast-EOM} or explicitly
\begin{equation}
\dot{\bL}_u=(\bD_0-u\bD_1)\times\bL_u=\bh_u\times\bL_u,
\end{equation}
where we use the compact notation $\bh_u\equiv\bD_0-u\bD_1$. This means in particular that
\begin{equation}\label{eq:Luplus-EOM}
\dot{L}_u^+=-i L_u^z h^+_u+i h_u^z L^+_u,
\end{equation}
where $L_u^\pm=L_u^x\pm i L_u^y$. Notice that in general both $L^x_u$ and $L^y_u$ are complex numbers.

Let us assume that initially all polarization vectors are aligned with the $z$ axis, except for a small perturbation $\delta\bD_v$ orthogonal to them. We denote by $\delta$ all perturbed quantities. If in the unperturbed situation we can find some $u$ such that the condition 
\begin{equation}\label{eq:Luz}
    L_u^z=0
\end{equation}
holds, then it follows that the first term on the right-hand side of Eq.~\eqref{eq:Luplus-EOM} drops out and the remaining motion is a rotation around the $z$ axis with frequency $h_u^z=D_0^z-u D_1^z$, i.e., the eigenfrequency is
$\Omega=D_0^z-u D_1^z$ that can be both real or complex. The condition Eq.~\eqref{eq:Luz} is recognized to be equivalent (in integral form) to the condition Eq.~\eqref{eq:fast-eigenvalue-linear}. It also follows that the corresponding eigenvector is aligned exactly with the Lax vector, namely that if the initial perturbation is chosen as
\begin{equation}
\delta \bD_v\propto \frac{v G_v}{v-u}(\mathbf{e}^x+i\mathbf{e}^y),
\end{equation}
where $\mathbf{e}^i$ is the unit vector along the axis $i$, it will evolve in time proportionally to itself with a frequency $\Omega=D^z_0-uD^z_1$. By replacing $u=(D^z_0-\Omega)/D^z_1$, we recover the eigenstates obtained by linear dispersion analysis in Eq.~\eqref{eq:eigenvectors-fast}.

Notice that this derivation closely parallels the derivation of the eigenfrequencies of the quantum Hamiltonian in Appendix~\ref{sec:app_diagonalization}.

\section{Three-flavor extension}

\label{sec:Three-Flavor}

The three-flavor case is notoriously more complicated, where the mean field of neutrino flavor is represented by $3\times3$ density matrices or by 8-dim Bloch vectors. Yet the technique of Lax vectors carries through in this case as well and one may identify similar Gaudin invariants and prove the integrability of the homogeneous EOMs.

\subsection{Bloch Vectors and Equations of Motion}

In the three-flavor case, the homogeneous flavor mean field is represented by $3\times3$ density matrices. For the axially symmetric fast flavor case each mode $v$ is represented by $\varrho_v$. It can be associated to an 8-dimensional Bloch vector using the decomposition \cite{Dasgupta:2007ws,Kimura:2003PhLA}
\begin{equation}
    \varrho_v=\sum_{i=1}^8 P^i_v \lambda_i,
\end{equation}
where $\lambda_i$ are the 8 Gell-Mann matrices. Here $\varrho_v$ is already written as a traceless matrix, because the trace of the matrix is the total number of neutrinos which is conserved in the absence of collisions. Therefore, one can still describe the dynamics in terms of vector equations of motion, although in 8-dimensional space. 

We recollect the main properties of the Gell-Mann matrices which will be useful in this discussion, namely
\begin{equation}
    [\lambda_i,\lambda_j]=2if_{ijk} \lambda_k,\quad
    \left\{\lambda_i,\lambda_j\right\}=\frac{4}{3}\delta_{ij} +2 d_{ijk} \lambda_k,
\end{equation}
where $f_{ijk}$ and $d_{ijk}$ are the structure constants. Therefore, the commutator of two matrices $\varrho_1$ and $\varrho_2$ can be compactly expressed as a vector product of their Bloch vectors $[\varrho_1,\varrho_2]=2i(\mathbf{P}_1\times \mathbf{P}_2)\cdot\pmb{\lambda}$, where
\begin{equation}
  (\mathbf{P}_1\times \mathbf{P}_2)_k = f_{ijk} P_1^i P_2^j.
\end{equation}
Notice that $f_{ijk}$ is antisymmetric under the exchange of any two indices.

We can now write the EOMs for the Bloch vectors of lepton number in an identical form as for two flavors
\begin{equation}\label{eq:eqsmotion3flavor}
    \dot\bD_v=(\mathbf{D}_0-v\mathbf{D}_1)\times\mathbf{D}_v.
\end{equation}
The antisymmetry of $f_{ijk}$ implies that this is once more a ``precession equation'' in the sense that $\bD_v$ moves in a direction orthogonal to $\bD_v$ so that its length is conserved.

This formal analogy implies that all of the properties we derived in the previous sections are still valid in the three-flavor case. For example, we can perform the exact same mapping that leads us to the equations of slow flavor conversions. Furthermore, it is still true that we can associate to any set of Bloch vectors obeying the EOMs~\eqref{eq:eqsmotion3flavor} a new set of Lax vectors obeying the same equations of motion. This means that a continuous set of Bloch vectors can still exhibit collective motions corresponding to a small number of degrees of freedom, e.g.\ a pendulum motion.

\subsection{How Many Degrees of Freedom?}

To investigate the integrability of the system, we need to identify both the number of dynamical degrees of freedom and the number of independent invariants.  In the two-flavor case, the number of Lagrangian degrees of freedom of $N$ Bloch vectors $\mathbf{D}_v$ is exactly $N$, since the length of each $\bD_v$ is conserved, and two angles are needed to describe the motion, one of which is the canonical conjugate of the other. In addition, the length of the Lax vector $\bL_v$ is also conserved, providing $N$ constants of the motion that are equivalent to the $N$ Gaudin invariants (Sec.~\ref{sec:HowMany}). Therefore, we had $N$ linearly independent invariants and the system was therefore integrable in the mechanical sense.

What is the number of degrees of freedom in the three-flavor system? To answer this question, we first need to identify the analogue of the length conservation for the three-dimensional Bloch vectors. The density matrices obey a commutator equation of the form
\begin{equation}
    \frac{\partial \varrho_v}{\partial t}=-i[\sH_v,\varrho_v].
\end{equation}
It is easily shown that the trace of $\varrho_v$ is conserved. As a consequence, the traces of the powers $\rho^n_v$ are conserved as well for any $n$. We usually consider $\varrho_v$ to denote the traceless part of the density matrix which can be expressed in terms of the Bloch vector $\bD_v$ and in this case for $n=1$, $\text{Tr}\,\rho_v=0$ identically. The non-trivial conservation laws come from $n=2$ and $n=3$, giving
\begin{equation}
    \bD_v^2=\bD_v\cdot\bD_v=\text{const.}
\end{equation}
and 
\begin{equation}
    \langle \mathbf{D}_v,\mathbf{D}_v,\mathbf{D}_v\rangle=\text{const.},
\end{equation}
where we define the triple product as $\langle A, B, C\rangle=d_{ijk} A_i B_j C_k$. For higher values of $n$, we find no new conservation laws. These two invariants of motion are the analogue of the fixed-length property of the two-flavor Bloch vectors. Therefore, for $N$ Bloch vectors with $8$ components each, only $6$ angles are needed to parameterize each of them. In analogy with the two-flavor case, we may say that the system admits $3N$ Lagrangian variables and $3N$ canonically conjugate variables.

Another way of looking at the number of variables comes from observing that the diagonal elements of the original density matrix of one mode of the neutrino radiation field are the usual occupation numbers, whereas the complex off-diagonal elements encode flavor coherence. In the two-flavor case, there are two diagonal elements and one independent off-diagonal one (one complex number or two real parameters). Altogether there is the trace, the length of the Bloch vector, and two angle variables. In the three-flavor case, there are three real diagonal elements and three independent complex off-diagonal ones (six real parameters). Altogether we have the trace, the length of the Bloch vector, and its triple product, and in addition six angle variables encoding flavor coherence information.

The physical interpretation of the conserved-length parameters follows if we imagine that initially all neutrinos begin in flavor eigenstates so that $\varrho={\rm diag}(f_{\nu_e},f_{\nu_\mu},f_{\nu_\tau})$ before making it traceless. We have already observed that the trace conservation implies that flavor evolution without collisions conserves $\sum_{\ell}f_{\nu_\ell}^n$  for any $n$. In the two-flavor case, the two conserved quantities are particle number $(f_{\nu_e}+f_{\nu_\mu})/2$ and the coefficient of $\sigma_3$ in the Pauli-matrix expansion, which is $(f_{\nu_e}-f_{\nu_\mu})/2$. Its absolute value is the length of the Bloch vector. In general, $(f_{\nu_e}-f_{\nu_\mu})/2$ is the $z$ component of the Bloch vector which is not conserved as flavor oscillations proceed. The
initial quantity $(f_{\nu_e}-f_{\nu_\mu})/2$ is the conserved length of the Bloch vector if neutrinos start in flavor eigenstates.

In the three-flavor case, the relevant parameters are the coefficients in the Gell-Mann matrix expansion, i.e., the eight components of the Bloch vector $\bP$, and in addition the coefficient $P_0=\frac{1}{3}{\rm Tr}\varrho$ of the unit matrix, which is not part of the Bloch vector. To represent the initial $\varrho$ that has only diagonal components, we only need
$P_0$ and $P_{3,8}$, the coefficients of the diagonal matrices $\lambda_{3,8}$, and find
\begin{subequations}
\begin{eqnarray}
    P_0&=&\frac{f_{\nu_e}+f_{\nu_\mu}+f_{\nu_\tau}}{3},
    \\
    P_3&=&\frac{f_{\nu_e}-f_{\nu_\mu}}{2},
    \\
    P_8&=&\frac{f_{\nu_e}+f_{\nu_\mu}-2f_{\nu_\tau}}{2\sqrt{3}}.
\end{eqnarray}    
\end{subequations}
The squared length of the Bloch vector $\bP\cdot\bP=
P_3^2+P_8^2=\frac{1}{3}(f_{\nu_e}^2+f_{\nu_\mu}^2+f_{\nu_\tau}^2-f_{\nu_e}f_{\nu_\mu}-f_{\nu_e}f_{\nu_\tau}-f_{\nu_\mu}f_{\nu_\tau})$ is conserved under flavor oscillations. The third conserved quantity is the triple product
$\langle \bP,\bP,\bP\rangle= \left(\sqrt{3} P_3^2 -P_8^2/\sqrt{3}\right)P_8=\frac{1}{18} (-f_{\nu_e}-f_{\nu_\mu}+2f_{\nu_\tau})(-f_{\nu_e}+2f_{\nu_\mu}-f_{\nu_\tau})(2f_{\nu_e}-f_{\nu_\mu}-f_{\nu_\tau})$.

\subsection{Gaudin Invariants and Integrability}

To identify the Gaudin invariants, it is more straightforward to express the Lax vectors in matrix form. The original EOMs are $\partial_t \sD_v=-i[\sH_v,\sD_v]$ 
with \edit{$\sH_v=\sD_0-v\sD_1$} or in vector form
$\partial_t \bD_v=\bH_v\times\bD_v$ with $\bH_v=\bD_0-v\bD_1$.
\edit{
In this case, it is convenient to start with the Lax vectors for an arbitrary complex parameter $u$, assumed to be different from any of the discrete values $v$ of the original system,
\begin{equation}
    \tilde{\mathbf{L}}_u=\sum_{v} \frac{\mathbf{D}_v v}{u-v}
    \quad\hbox{and}\quad
    \tilde{\sL}_u=\tilde{\mathbf{L}}_u\cdot\pmb{\lambda},
\end{equation}
obeying the original EOMs $\partial_t \tilde{\bL}_v=\bH_v\times \tilde{\bL}_v$ or in matrix form $\partial_t \tilde{\sL}_v=-i[\sH_v,\tilde{\sL}_v]$. 
Since the motion is an instantaneous precession, for any $u$, \smash{$\mathrm{Tr}(\tilde{\sL}_u^n)$} is conserved for any $n$. As in the case of $\mathrm{Tr}(\sD_u^n)$, we can find independent and non-trivial invariants only for $n=2$ or~3. Hence, we have a continuum of invariants for any complex value of $u$, which clearly are not all independent. To identify the analog of the Gaudin invariants, we are going to take the limit for $u\to \bar{v}$, where $\bar{v}$ is one of the discrete velocities. This limit must be taken with care. Let us separate the regular part from the pole term
\begin{equation}
    \tilde{\bL}_u= \frac{\bar{v}\bD_{\bar{v}}}{u-\bar{v}}+\bL_{u},
\end{equation}
with
\begin{equation}
    \bL_{u}=\sum_{v\neq \bar{v}}\frac{\bD_v v}{u-v}.
\end{equation}
We will need only an expansion in $u-\bar{v}$ up to the first order
\begin{equation}
    \tilde{\bL}_u\simeq\frac{\bar{v}\bD_{\bar{v}}}{u-\bar{v}}+\bL_{\bar{v}}-\bM_{\bar{v}}(u-\bar{v}),
\end{equation}
where
\begin{equation}
    \bM_{\bar{v}}=\sum_{v\neq \bar{v}}\frac{v\bD_v}{(\bar{v}-v)^2}.
\end{equation}
The analogous matrices are
\begin{equation}
    \tilde{\sL}_u\simeq \sL_{\bar{v}}+\frac{\sD_{\bar{v}}\bar{v}}{u-\bar{v}}-\sM_{\bar{v}}(u-\bar{v}).
\end{equation}

We can now deduce conservation laws for the vectors $\bL_{\bar{v}}$ using the conservation of the pole terms of $\mathrm{Tr}(\tilde{\sL}_u^n)$ for $u\to \bar{v}$. For $n=2$, the pole contributions are
\begin{equation}
    \mathrm{Tr}(\tilde{\sL}_u^2)\simeq\frac{\mathrm{Tr}(\sD_{\bar{v}}^2)\bar{v}^2}{(u-\bar{v})^2}+\frac{2\bar{v}\mathrm{Tr}(\sL_{\bar{v}}\sD_{\bar{v}})}{u-\bar{v}};
\end{equation}
we do not write the subdominant regular terms, since they would require more precision in the expansion of $\tilde{\sL}_u$. Hence, we immediately find the conservation of $|\bD_v|^2$, which we already knew about, and the conservation of $I_v=\bL_v\cdot \bD_v$. We now turn to $n=3$
\begin{eqnarray}
    \mathrm{Tr}\bigl(\tilde{\sL}_u^3\bigr)&\simeq& \frac{\mathrm{Tr}\bigl(\sD_{\bar{v}}^3\bigr)\bar{v}^3}{(u-\bar{v})^3}+\frac{3\bar{v}^2\mathrm{Tr}\bigl(\sL_{\bar{v}} \sD_{\bar{v}}^2\bigr)}{(u-\bar{v})^2}\\ \nonumber &+&\frac{3\bar{v}}{u-\bar{v}}\left[\mathrm{Tr}(\sL_{\bar{v}}^2 \sD_{\bar{v}})-\bar{v}\mathrm{Tr}(\sD_{\bar{v}}^2 \sM_{\bar{v}})\right].
\end{eqnarray}
From here we see that there are two families of additional invariants 
\begin{equation}
    J_v=\langle \bL_v, \bD_v,\bD_v\rangle
\end{equation}
and
\begin{equation}\label{eq:Kinvariant}
   K_v=\langle \bL_v, \bL_v,\bD_v\rangle-v\langle\bD_v,\bD_v,\bM_v\rangle.
\end{equation}
These are the generalizations of the two-flavor Gaudin invariants to three flavors. Notice that these invariants are all linearly independent of one another. Therefore, we have $3N$ Lagrangian degrees of freedom, and $3N$ associated conservation laws for $I_v$, $J_v$, and $K_v$. This guarantees integrability in the mechanical sense also for the three-flavor motion.}

\section{Conclusions}

\label{sec:Conclusion}

The homogeneous nonlinear equations of motion for collective neutrino flavor oscillations have two often-studied limits, the single-angle case of slow oscillations and fast oscillations, corresponding to the multi-angle case in the limit of vanishing neutrino masses. In the classical limit (mean-field approach), both cases correspond to ensembles of classical interacting spins, but with different interaction structure.

Whereas the ``slow Hamiltonian'' corresponds to well-studied cases in other fields, in particular the BCS Hamiltonian in the theory of superconductivity, the ``fast Hamiltonian'' appears to be less prominent. On the other hand, both cases have many similarities. It has been previously observed that both cases have soliton (pendulum-like) solutions, the nonlinear manifestation of the run-away solutions of a linear mode analysis.

We have systematically juxtaposed the two cases and developed the similarities and differences between them. Our starting point is the observation that the EOMs of the fast system can always be mapped on an abstract slow system. For the special case when the fast system has a pendulum solution, our transformation has the intuitive interpretation that, from the perspective of the moving pendulum, it is the ``direction of gravity'' the moves like a ``reciprocal'' pendulum with EOMs formally equivalent to a slow system. While transforming the EOMs to co-rotating frames is a standard procedure in this field, probably this is the first time one has used a more general transformation.

On the other hand, a slow system can be mapped on an equivalent fast system only if its system of spins (or Bloch vectors) obeys an additional constraint. This fundamental difference between the two cases is traced to the presence of an external vector in the slow system, the mass direction $\bB$ in flavor space, whereas the fast system has no preferred direction except for the conserved total angular momentum, which is determined by initial conditions, not the Hamiltonian itself.

We have adapted the powerful tool of Lax vectors, introduced in this field in studies of the BCS Hamiltonian, to the fast system. In this way, it is straightforward to identify the invariants of the systems and especially the Gaudin invariants, but also offers another perspective on the normal mode analysis, allowing one to derive the dispersion relation in a single line. 

In the continuum (thermodynamic) limit, this dispersion relation covers only the collective modes, whereas the non-collective ones require a different interpretation. Using the equivalence of our linearized system to the Vlasov equation, we highlight the equivalence of the non-collective modes to the well-known Case-Van Kampen modes. In this way, decoherence of a generic initial condition is a natural feature of our system.

The Gaudin invariants immediately prove the integrability of the system. We explicitly show that the integrability also translates to the quantum case.

These techniques and many of our conclusions carry over to the three-flavor case, where the Bloch vectors and Lax vectors live in an eight-dimensional space. While part of this extension is straightforward, it involves a number of nontrivial steps and new invariants, owing to the SU(3) structure of this problem.

Relative to the question of neutrino flavor evolution in real astrophysical environments, our homogeneous system is probably too constrained to serve as a realistic proxy. On the other hand, it provides surprising insights on the foundational properties of the nonlinear EOMs and their possible solutions.

\edit{\section*{Note added after publication}
Returning to the subject of mechanical spin systems three years after publication of this paper, we have noticed a number of technical inconsistencies in the context of identifying the Gaudin invariants. The conclusions for the two-flavor system remain unchanged. For the three-flavor system, the third class of Gaudin invariants, now shown as Eq.~\eqref{eq:Kinvariant}, is actually different from the earlier version. All changes relative to the published version are shown in blue.
}

\section*{Acknowledgements}

GR acknowledges support by the German Research Foundation (DFG) through the Collaborative Research Centre ``Neutrinos and Dark Matter in Astro and Particle Physics (NDM),'' Grant SFB-1258, and under Germany's Excellence Strategy through the Cluster of Excellence ORIGINS EXC-2094-390783311. DFGF is supported by the Villum Fonden under Project No.\ 29388 and the European Union's Horizon 2020 Research and Innovation Program under the Marie Sk{\l}odowska-Curie Grant Agreement No.\ 847523 ``INTERACTIONS.''

\appendix

\section{Comoving frames}
\label{app:transformation}

Precession equations of the type used throughout this paper can become more transparent in a moving frame in flavor space. Frames rotating around the $z$-direction~\cite{Duan:2005cp} have become standard. However, one has to be careful to change everything consistently. In the fast EOM \eqref{eq:EOM-fast-A} we may simply leave out $\bD_0$ by going to a frame rotating around the fixed $\bD_0$ with unit frequency. On the other hand, going to a frame co-rotating with frequency $w_{\rm c}$ around $\bB$ in the slow EOM \eqref{eq:EOM-slow-A} changes $w\to w-w_{\rm c}$.

To derive the formal transformation between two frames we consider the precession equations
\begin{equation}\label{eq:EOM-Pw}
    \dot\bP_w=\bJ_w\times\bP_w,
\end{equation}
where $\bP_w$ is a Bloch vector that depends on some attribute $w$ and precesses around some other vector $\bJ_w$ that may itself depend on time. Moreover, we consider another precession equation of the form
\begin{equation}
    \dot\bR=\bH\times\bR
\end{equation}
and we want to express the former in a frame that comoves with $\bR$.

The formal transformation is more easily derived using instead the commutator form of the EOMs. We recall that any $2\times2$ Hermitean matrix $\sP$ can be expressed as $\sP=\frac{1}{2}({\rm Tr}\,\sP+\bP\cdot{\bm\sigma})$ in terms of the Pauli matrices and a Bloch vector $\bP$. The commutation relations of the Pauli matrices imply that an EOM of the form 
\begin{equation}\label{eq:EOM-commutator}
    i\partial_t\sP=[\sH,\sP]
\end{equation}
is represented by ${\rm Tr}\,\sP=0$ and $\dot\bP=\bH\times\bP$. The EOM~\eqref{eq:EOM-commutator} is formally solved by $\sR_t=\sU_t\sR_0\sU^\dagger_t$ with
\begin{equation}\label{eq:U-matrix}
\sU_t={\cal T} \exp\left[-i\int_0^t\sH_{t'}dt'\right],
\end{equation}
where ${\cal T}$ is the time-ordering operator. This transformation provides the compound effect of infinitesimal precessions around $\bH_t$ that is itself moving. The content of Eq.~\eqref{eq:U-matrix} can also be expressed as
\begin{equation}
    i\partial_t\sU_t=\sH_t\sU_t
    \quad\hbox{and}\quad
    i\partial_t\sU_t^\dagger=-\sU_t^\dagger\sH_t,
\end{equation}
where it was assumed that $\sH$ is a Hermitean matrix. One can verify that indeed $i\partial_t\sR=i\partial_t (\sU\sR_0\sU^\dagger)=\sH \sU\sR_0\sU^\dagger-\sU\sR_0\sU^\dagger\sH=[\sH,\sR]$.

We now denote matrices in the new frame with a tilde, i.e., $\tilde\sP_w=\sU^\dagger\sP_w\sU$. In particular, $\tilde\sR=\sU^\dagger\sR\sU=\sU^\dagger\sU R_0\sU^\dagger\sU=\sR_0$. So indeed $\sR=\sR_0$ at all times, which was the original goal of the transformation. The equivalent of Eq.~\eqref{eq:EOM-Pw} is 
$i\partial_t\sP_w=[\sJ_w,\sP_w]$. In the new frame it is explicitly 
$i\partial_t\tilde\sP_w=i\partial_t(\sU^\dagger\sP_w\sU)
=(i\partial_t\sU^\dagger)\sP_w\sU+\sU^\dagger\sP_w(i\partial_t\sU)+\sU^\dagger(i\partial_t\sP_w)\sU
=-\sU^\dagger\sH\sP_w\sU+\sU^\dagger\sP_w\sH\sU+\sU^\dagger[\sJ_w,\sP_w]\sU
=-[\tilde\sH,\tilde\sP_w]+[\tilde\sJ_w,\tilde\sP_w]$ and finally
\begin{equation}
    i\partial_t\tilde\sP_w=\bigl[\,\tilde\sJ_w-\tilde\sH,\tilde\sP_w\bigr].
\end{equation}
In this form, our result also applies to three flavors. Returning to Bloch vectors, the result is equivalent to
\begin{equation}
  \partial_t\tilde\bP_w=\bigl(\,\tilde\bJ_w-\tilde\bH\bigr)\times\tilde\bP_w.
\end{equation}
This is the same prescription that was argued in the main text: To take out the motion engendered by $\bH$ we must subtract $\tilde\bH$ in the transformed frame from the vector $\tilde\bJ_w$ that generates the original motion.

\section{Linear stability analysis in the thermodynamic limit}
\label{sec:thermodynamic_limit}

The dispersion relation Eq~\eqref{eq:slow-eigenvalue} is unsatisfactory from a mathematical standpoint, because it involves an integral which is singular for real frequencies. A prescription on how to deal with this singularity should be given. This issue has historically first been solved in the context of collisionless plasma oscillations. Indeed, Eq.~\eqref{eq:schrodinger} is mathematically identical to the linearized Vlasov equation describing the evolution of an initial spatially monochromatic small perturbation $\delta f(\mathbf{r},\mathbf{v})=\delta f_\mathbf{k}(\mathbf{v}) e^{i\mathbf{k}\cdot\mathbf{r}}$ in the distribution function of electrons in a plasma (here $\mathbf{v}$ is the electron speed)\footnote{We use here unrationalized units (fine-structure constant $\alpha=e^2$) that are conventionally applied in plasma physics, but here put the speed of light $c=1$.}
\begin{equation}\label{eq:vlasov}
\frac{\partial \delta f_\mathbf{k}}{\partial t}+i\mathbf{k}\cdot\mathbf{v}\delta f_\mathbf{k}=-\frac{4\pi i e^2}{m k^2}\mathbf{k}\cdot\frac{\partial f_0}{\partial\mathbf{v}}\int \delta f_\mathbf{k}\,d^3 \mathbf{v},
\end{equation}
where $f_0(\mathbf{v})$ is the unperturbed distribution, $e$ and $m$ are the electron charge and mass; see, e.g., Ref.~\cite{Landau:1946jc}. Let $\mathbf{k}$ be aligned with the $z$ axis. We now define
\begin{equation}
P_\mathbf{k}=\int \delta f_\mathbf{k} dv_x dv_y,
\end{equation}
and similarly
\begin{equation}
g=-\frac{4\pi e^2}{mk}\int \frac{\partial f_0}{\partial v_z} dv_x dv_y.
\end{equation}
Integrating Eq.~\eqref{eq:vlasov} over $v_x$ and $v_y$, we obtain
\begin{equation}
i\frac{\partial P_\mathbf{k}(v_z)}{\partial t}-kv_zP_\mathbf{k}(v_z)=-g\int P_\mathbf{k}(v_z) dv_z.
\end{equation}
This equation is indeed identical to Eq.~\eqref{eq:schrodinger}, with the identification $k v_z=w$. The singularities in the dispersion relation appear in the same way in this analogous problem; below, we summarize the main results from the literature concerning this issue.

The eigenmodes of the linearized problem can be separated into collective and non-collective modes, as is done in Ref.~\cite{Capozzi:2019lso}. The collective modes \smash{$p^{xy}_{w,i}$} appear as discrete complex solutions $\Omega_i$ of the dispersion relation~\eqref{eq:slow-eigenvalue}, and, as shown in the text, they always appear in pairs of complex conjugate solutions. The number of such collective modes depend on the spectrum $g_w$; for example, for non-crossed spectra there are no collective modes.

The non-collective modes are not predicted by the dispersion relation~\eqref{eq:slow-eigenvalue}. Rather, they correspond to a continuum of modes with real frequencies $\Omega_v=v B^z-P_0^z$, and do not show up in our dispersion relation because their eigenfunctions are singular
\begin{equation}\label{eq:singular_eigenmodes}
p^{xy}_{w,v}\propto \delta(w-v)-\frac{g_w}{\Omega_v+P^z_0-w B^z}\frac{1}{1+f_v},
\end{equation}
where
\begin{equation}
f_v=\int_{\mathrm{PV}}\frac{g_w}{\Omega_v+P_0^z-wB^z} dw.
\end{equation}
The pole in the eigenmode~\eqref{eq:singular_eigenmodes} is meant to be integrated in principal value (PV). 

By explicit substitution, one can verify that these eigenmodes indeed satisfy Eq.~\eqref{eq:schrodinger}. These modes were first identified as the so-called Case-Van Kampen modes \cite{VanKampen:1955wh} in the case of collisionless plasma oscillations, where it was shown that the collective and non-collective modes together form a complete set of eigenmodes for the linearized Vlasov equation. By the same token, in our problem of neutrino oscillation, any initial perturbation $P^{xy}_w(t=0)$ can be decomposed in collective and non-collective eigenmodes which evolve independently, so that the full evolution can be written (see, e.g., Ref.~\cite{Sagan:1993es})
\begin{equation}\label{eq:evolution_pol_vectors}
P^{xy}_w(t)=\sum_i a_i p^{xy}_{w,i} e^{-i \Omega_i t}+\int dv b(v) p^{xy}_{w,v} e^{-i\Omega_v t},
\end{equation}
where the coefficients $a_i$ and $b(v)$ are determined by the initial conditions.

However, the integrated lepton number $P^{xy}(t)=\int dw P^{xy}_w(t)$ has an entirely different dynamics in the continuum limit. The reason is that the superposition of the Case-Van Kampen modes, evolving with a continuum of frequencies in time, translates into a damped behavior. This behavior is the exact analog of Landau damping in collective plasma oscillations; even though the individual polarization vectors have non-decaying terms in their perturbation, because of their fast precession with incommensurate frequencies they lead to a perturbation in the integrated lepton number that decays in time. 

The dynamics of $P^{xy}(t)$ could be obtained by explicitly integrating Eq.~\eqref{eq:evolution_pol_vectors}. However, it is easier to circumvent the expansion in eigenstates and directly look at the evolution in time of the initial condition using the Laplace method, following the strategy in Ref.~\cite{Landau:1946jc}. By this method, one finds that $P^{xy}(t)$ evolves in time as
\begin{equation}
P^{xy}(t)=\sum_\alpha A_\alpha e^{-i \tilde{\Omega}_\alpha t},
\end{equation}
where $A_\alpha$ are determined by the initial condition and $\tilde{\Omega}_\alpha$ are the solutions of the modified dispersion relation
\begin{equation}
\int_C dw \frac{g_w}{wB^z-P_0^z-\tilde{\Omega}_\alpha}=1,
\end{equation}
where the contour of integration $C$ is chosen as the real axis if $\mathrm{Im}(\tilde{\Omega}_\alpha)>0$, therefore identical to Eq.~\eqref{eq:slow-eigenvalue}, whereas for $\mathrm{Im}(\tilde{\Omega}_\alpha)<0$ the contour of integration is deformed so as to pass below the pole $w=(\tilde{\Omega}_\alpha+P^z_0)/B^z$. 

The difference in the contour definition for growing ($\mathrm{Im}\,\tilde{\Omega}_\alpha>0$) and shrinking modes ($\mathrm{Im}\,\tilde{\Omega}_\alpha<0$) originates from causality requirements. Because of this difference, the frequencies $\tilde{\Omega}_\alpha$ do not appear in complex conjugate pairs. The modified dispersion relation admits new shrinking modes without the corresponding growing ones. Physically, these shrinking modes come from Landau damping of the initial perturbation, due to the phase mixing of the many polarization vectors with incommensurate frequencies. The symmetry between growing and shrinking modes is therefore broken by the second principle of thermodynamics, which implies the decoherence of the motion of individual polarization vectors into a damping of the integrated lepton number.

\section{Diagonalization of the quantum fast flavor system}
\label{sec:app_diagonalization}

The quantum version of the fast flavor system can be regarded as a system of spin-$1/2$ $\bS_v$ with the Hamiltonian
\begin{equation}\label{eq:full_hamiltonian}
\mathcal{H}=\frac{\mu}{2}\sum_{v,v'}(1-vv')\,\bS_v \cdot \bS_{v'}.
\end{equation}
We assume for definiteness a discrete system of $N$ spins. Here we have assumed that each mode with cosine of zenith angle $v$ is occupied by exactly one neutrino; the method is easily generalizable to the case in which there is a varying occupation number. This Hamiltonian is solvable in the quantum mechanical sense. A first indication of this comes from the fact that the Gaudin invariants, which we showed to be conserved in the analog classical system, are also conserved in the quantum sense, meaning that their operators
\begin{equation}
I_v=\sum_{v'\neq v}\frac{v'\,\bS_{v'}\cdot\bS_v}{v-v'}
\end{equation}
commute with the Hamiltonian, as can be verified by explicit computation.

The existence of the Gaudin invariants implies that this Hamiltonian can be exactly diagonalized. In this appendix, we show how the eigenstates of the Hamiltonian can be constructed. Since the fast flavor system is so similar to the slow flavor system, the diagonalization procedure is very similar to the one shown in Ref.~\cite{Pehlivan:2011hp}: whenever necessary, we will emphasize the differences.

A first, fundamental difference to the slow flavor system is the following. The Hamiltonian~\eqref{eq:full_hamiltonian} is invariant under a simultaneous rotation of all polarization vectors. This is not the case for the slow system, where there is a privileged direction fixed by the $\bB$ vector. The rotational invariance of the fast flavor system implies the conservation of the generator of collective rotation
\begin{equation}
\bD_0=\sum_v \bS_v,
\end{equation}
which indeed commutes with the Hamiltonian. Therefore, we reach again the conclusion stated multiple times in the main text that the fast flavor system is analogous to a slow flavor system endowed with an exactly conserved vector. For the fast flavor system, such a vector is $\bD_0$, which is conserved by rotational invariance.

Let us now determine the eigenstates. The Hamiltonian~\eqref{eq:full_hamiltonian} admits two trivial eigenstates, in which all the spins are aligned along, say, the $z$ direction. We will use the notation $\ket{S,S_z}$ to denote the eigenstates of the total spin with eigenvalues $S$ and $z$ component $S_z$. Then the two eigenstates identified above are $\ket{\frac{N}{2},\pm\frac{N}{2}}$. However, because of rotational invariance, there is nothing special about the $z$ direction, and these two states rotated in an arbitrary way are still eigenstates of the Hamiltonian with the same eigenvalues. Therefore, the entire multiplet of $2N+1$ states $\ket{\frac{N}{2},S_z}$, with $S_z$ ranging from $-\frac{N}{2}$ to $\frac{N}{2}$, are degenerate eigenstates of the Hamiltonian. This stands in sharp contrast to a generic slow flavor system, for which only the two eigenstates $\ket{\frac{N}{2},\pm\frac{N}{2}}$ with spin totally aligned to the $z$ axis defined by $\bB$ are eigenstates.

We now need to identify the remaining eigenstates. We will classify them in terms of their $z$ component of spin $S_z$, and we will start looking for eigenstates containing all spins aligned, for example negatively, along the $z$ axis, except for a single spin flipped (therefore, $S_z=-\frac{N}{2}+1$). If there were no interaction, all $N$ states of the form $\ket{\uparrow\downarrow...\downarrow}$, $\ket{\downarrow\uparrow...\downarrow}$,..., $\ket{\downarrow\downarrow...\uparrow}$ would be degenerate eigenstates of the Hamiltonian. The presence of the interaction splits the degeneracy and selects $N$ combinations of these states, which we now identify; these states are essentially the analog of the magnon excitations in the one-dimensional Ising model, with the difference that our Hamiltonian~\eqref{eq:full_hamiltonian} breaks translational invariance, so that the form of the eigenstates is not as simple as in the Ising model.

To identify the form of the eigenstates with a single spin flip, we use the Bethe ansatz and assume that they can be written as
\begin{equation}
\ket{\psi_u}=\mathcal{Q}_u\ket{{\textstyle\frac{N}{2},-\frac{N}{2}}},
\end{equation}
where $\mathcal{Q}_u$ is an operator we are now going to identify. In order to do so, let us recall that the Lax vectors introduced in the main text
\begin{equation}
\bL_u=\sum_v\frac{v\bS_v}{u-v}
\end{equation}
follow a pure precession motion
\begin{equation}
\dot{\bL}_u=(\bD_0-u\bD_1)\times\bL_u=i\left[\mathcal{H},\bL_u\right].
\end{equation}
Defining the vector $\bh_u=\bD_0-u\bD_1$ implies that
\begin{equation}
\left[\mathcal{H},L_u^+\right]= L_u^+ h_u^z-h_u^+ L_u^z,
\end{equation}
where $L_u^\pm=L_u^x\pm i L_u^y$ and so forth.

We now make the ansatz that the operator $\mathcal{Q}_u$ introduced above coincides with $L_u^+$, and determine how the Hamiltonian acts on the state $\ket{\psi_u}$
\begin{equation}
\mathcal{H}\ket{\psi_u}=[\mathcal{H},L_u^+]\ket{{\textstyle\frac{N}{2},-\frac{N}{2}}}+L_u^+\mathcal{H}\ket{{\textstyle\frac{N}{2},-\frac{N}{2}}}.
\end{equation}
The last term is $\mathcal{H}\ket{\frac{N}{2},-\frac{N}{2}}=E_0\ket{\frac{N}{2},-\frac{N}{2}}$, where $E_0$ is the ground state energy. In the first term, we notice that $L^+_u h_z^u \ket{\frac{N}{2},-\frac{N}{2}}=e_u\ket{\psi_u}$, where $e_u$ is the eigenvalue of the operator $h_z^u$ acting on the ground state $\ket{\frac{N}{2},-\frac{N}{2}}$, namely
\begin{equation}
e_u=-\sum_v \frac{1-uv}{2}.
\end{equation}
Therefore, we obtain
\begin{equation}
\mathcal{H}\ket{\psi_u}=\left(E_0+e_u\right)\ket{\psi_u}-h^+_u L^z_u \ket{{\textstyle\frac{N}{2},-\frac{N}{2}}}.
\end{equation}
Therefore, if $u$ is chosen as one of the roots of $L^z_u \ket{\frac{N}{2},-\frac{N}{2}}$, namely one of the roots of the Bethe equation
\begin{equation}
\sum_v \frac{v}{u-v}=0,
\end{equation}
the last term vanishes and the state $\ket{\psi_u}$ is indeed an eigenstate of the Hamiltonian. The Bethe equation admits therefore $N-1$ eigenstates with a single spin flip; an additional such eigenstate is contained among the multiplet of $2N+1$ eigenstates of the total spin $D_0^z$, namely the state $\ket{\frac{N}{2},-\frac{N}{2}+1}$.

This procedure can be generalized to states with more than one spin flip. For example, one can prove that the commutator
\begin{eqnarray}
&&[\mathcal{H},L^+_u L^+_v]=L^+_u L^+_v(h^z_u+h^z_v+1-uv) \\ \nonumber &&-L^+_u h^+_v \left(L^z_v-\frac{u}{v-u}\right)-L^+_v h^+_u \left(L^z_v-\frac{v}{u-v}\right).
\end{eqnarray}
Therefore, the state $L^+_u L^+_v \ket{\frac{N}{2},-\frac{N}{2}}$ is an eigenstate of the Hamiltonian provided that the last two terms in parenthesis vanish identically, namely if the Bethe equations are satisfied
\begin{equation}
J^z_v=\frac{u}{v-u},\;J^z_u=\frac{v}{u-v},
\end{equation}
with the compact notation $L^z_u \ket{\frac{N}{2},-\frac{N}{2}}=J^z_u \ket{\frac{N}{2},-\frac{N}{2}}$. The eigenvalue of the Hamiltonian on this new eigenstate is $E_0+e_u+e_v+1-uv$.

Therefore, the repeated application of the operator $L^+_{u_1}...L^+_{u_n}$ on the ground state generates a novel eigenstate with eigenvalue $E_0+\sum_i e_{u_i}+\frac{1}{2}\sum_{i,j\neq i} (1-u_i u_j)$, provided that the Bethe equations are satisfied
\begin{equation}
J^z_{u_i}=\sum_j \frac{u_j}{u_i-u_j},
\end{equation}
which completes our proof.

\bibliographystyle{bibi}
\bibliography{Biblio}

@article{Chakraborty:2016yeg,
    author = "Chakraborty, Sovan and Hansen, Rasmus and Izaguirre, Ignacio and Raffelt, Georg",
    title = "{Collective neutrino flavor conversion: Recent developments}",
    eprint = "1602.02766",
    archivePrefix = "arXiv",
    primaryClass = "hep-ph",
    doi = "10.1016/j.nuclphysb.2016.02.012",
    journal = "Nucl. Phys. B",
    volume = "908",
    pages = "366--381",
    year = "2016"
}

@article{Anderson:1958zza,
    author = "Anderson, P. W.",
    title = "{Random-phase approximation in the theory of superconductivity}",
    doi = "10.1103/PhysRev.112.1900",
    journal = "Phys. Rev.",
    volume = "112",
    pages = "1900--1916",
    year = "1958"
}

@article{Sagan:1993es,
    author = "Sagan, David",
    title = "{On the physics of Landau damping}",
    doi = "10.1119/1.17547",
    journal = "Am. J. Phys. ",
     volume = "62",
    pages = "450",
    year = "1994"
}

@article{Landau:1946jc,
    author = "Landau, L. D.",
    title = "{On the vibrations of the electronic plasma}",
    journal = "J. Phys. (USSR)",
    volume = "10",
    pages = "25--34",
    year = "1946"
}

@article{VanKampen:1955wh,
    author = "Van Kampen, N. G.",
    title = "{On the theory of stationary waves in plasmas}",
    doi = "10.1016/S0031-8914(55)93068-8",
    journal = "Physica",
    volume = "21",
    pages = "949--963",
    year = "1955"
}

@article{Dasgupta:2007ws,
    author = "Dasgupta, Basudeb and Dighe, Amol",
    title = "{Collective three-flavor oscillations of supernova neutrinos}",
    eprint = "0712.3798",
    archivePrefix = "arXiv",
    primaryClass = "hep-ph",
    reportNumber = "TIFR-TH-07-36",
    doi = "10.1103/PhysRevD.77.113002",
    journal = "Phys. Rev. D",
    volume = "77",
    pages = "113002",
    year = "2008"
}

@ARTICLE{Kimura:2003PhLA,
       author = {{Kimura}, Gen},
        title = "{The Bloch vector for N-level systems}",
      journal = {Phys. Lett. A},
         year = 2003,
        month = aug,
       volume = {314},
       number = {5-6},
        pages = {339-349},
          doi = {10.1016/S0375-9601(03)00941-1},
archivePrefix = {arXiv},
       eprint = {quant-ph/0301152},
 primaryClass = {quant-ph},
       adsurl = {https://ui.adsabs.harvard.edu/abs/2003PhLA..314..339K}
}

@article{Capozzi:2019lso,
    author = "Capozzi, Francesco and Raffelt, Georg and Stirner, Tobias",
    title = "{Fast Neutrino Flavor Conversion: Collective Motion vs. Decoherence}",
    eprint = "1906.08794",
    archivePrefix = "arXiv",
    primaryClass = "hep-ph",
    reportNumber = "MPP-2019-120",
    doi = "10.1088/1475-7516/2019/09/002",
    journal = "JCAP",
    volume = "09",
    pages = "002",
    year = "2019"
}

@ARTICLE{Barkov:2004a,
       author = {{Barankov}, R.~A. and {Levitov}, L.~S. and {Spivak}, B.~Z.},
        title = "{Collective Rabi Oscillations and Solitons in a Time-Dependent BCS Pairing Problem}",
      journal = {Phys. Rev. Lett.},
         year = 2004,
        month = oct,
       volume = {93},
       number = {16},
          eid = {160401},
        pages = {160401},
          doi = {10.1103/PhysRevLett.93.160401},
archivePrefix = {arXiv},
       eprint = {cond-mat/0312053},
 primaryClass = {cond-mat.mes-hall},
       adsurl = {https://ui.adsabs.harvard.edu/abs/2004PhRvL..93p0401B}
}

@ARTICLE{Barkov:2004b,
       author = {{Barankov}, R.~A. and {Levitov}, L.~S.},
        title = "{Atom-Molecule Coexistence and Collective Dynamics Near a Feshbach Resonance of Cold Fermions}",
      journal = {Phys. Rev. Lett.},
         year = 2004,
        month = sep,
       volume = {93},
       number = {13},
          eid = {130403},
        pages = {130403},
          doi = {10.1103/PhysRevLett.93.130403},
archivePrefix = {arXiv},
       eprint = {cond-mat/0405178},
       adsurl = {https://ui.adsabs.harvard.edu/abs/2004PhRvL..93m0403B}
}

@ARTICLE{Patwardhan:2023,
       author = {{Patwardhan}, Amol V. and {Cervia}, Michael J. and {Rrapaj}, Ermal and {Siwach}, Pooja and {Balantekin}, A.~B.},
        title = "{Many-body collective neutrino oscillations: recent developments}",
         year = 2022,
        month = dec,
          eid = {arXiv:2301.00342},
        pages = {arXiv:2301.00342},
archivePrefix = {arXiv},
       eprint = {2301.00342},
 primaryClass = {hep-ph},
       adsurl = {https://ui.adsabs.harvard.edu/abs/2023arXiv230100342P}
}

@article{Izaguirre:2016gsx,
    author = "Izaguirre, Ignacio and Raffelt, Georg and Tamborra, Irene",
    title = "{Fast Pairwise Conversion of Supernova Neutrinos: A Dispersion-Relation Approach}",
    eprint = "1610.01612",
    archivePrefix = "arXiv",
    primaryClass = "hep-ph",
    reportNumber = "INT-PUB-16-023, MPP-2016-266",
    doi = "10.1103/PhysRevLett.118.021101",
    journal = "Phys. Rev. Lett.",
    volume = "118",
    number = "2",
    pages = "021101",
    year = "2017"
}

@article{Banerjee:2011fj,
    author = "Banerjee, Arka and Dighe, Amol and Raffelt, Georg",
    title = "{Linearized flavor-stability analysis of dense neutrino streams}",
    eprint = "1107.2308",
    archivePrefix = "arXiv",
    primaryClass = "hep-ph",
    reportNumber = "MPP-2011-81, TIFR-TH-11-30",
    doi = "10.1103/PhysRevD.84.053013",
    journal = "Phys. Rev. D",
    volume = "84",
    pages = "053013",
    year = "2011"
}

@article{Morinaga:2021vmc,
    author = "Morinaga, Taiki",
    title = "{Fast neutrino flavor instability and neutrino flavor lepton number crossings}",
    eprint = "2103.15267",
    archivePrefix = "arXiv",
    primaryClass = "hep-ph",
    doi = "10.1103/PhysRevD.105.L101301",
    journal = "Phys. Rev. D",
    volume = "105",
    number = "10",
    pages = "L101301",
    year = "2022"
}

@article{Dasgupta:2021gfs,
    author = "Dasgupta, Basudeb",
    title = "{Collective Neutrino Flavor Instability Requires a Crossing}",
    eprint = "2110.00192",
    archivePrefix = "arXiv",
    primaryClass = "hep-ph",
    reportNumber = "TIFR/TH/21-15",
    doi = "10.1103/PhysRevLett.128.081102",
    journal = "Phys. Rev. Lett.",
    volume = "128",
    number = "8",
    pages = "081102",
    year = "2022"
}

@article{Dasgupta:2009mg,
    author = "Dasgupta, Basudeb and Dighe, Amol and Raffelt, Georg G. and Smirnov, Alexei {\relax Yu}.",
    title = "{Multiple Spectral Splits of Supernova Neutrinos}",
    eprint = "0904.3542",
    archivePrefix = "arXiv",
    primaryClass = "hep-ph",
    reportNumber = "MPP-2009-33",
    doi = "10.1103/PhysRevLett.103.051105",
    journal = "Phys. Rev. Lett.",
    volume = "103",
    pages = "051105",
    year = "2009"
}

@article{Samuel:1993uw,
    author = "Samuel, Stuart",
    title = "{Neutrino oscillations in dense neutrino gases}",
    reportNumber = "IUHET-244",
    doi = "10.1103/PhysRevD.48.1462",
    journal = "Phys. Rev. D",
    volume = "48",
    pages = "1462--1477",
    year = "1993"
}

@article{Samuel:1995ri,
    author = "Samuel, Stuart",
    title = "{Bimodal coherence in dense selfinteracting neutrino gases}",
    eprint = "hep-ph/9604341",
    archivePrefix = "arXiv",
    reportNumber = "MPI-PHT-95-57, CCNY-HEP-95-5",
    doi = "10.1103/PhysRevD.53.5382",
    journal = "Phys. Rev. D",
    volume = "53",
    pages = "5382--5393",
    year = "1996"
}

@article{Pantaleone:1992eq,
    author = "Pantaleone, James T.",
    title = "{Neutrino oscillations at high densities}",
    reportNumber = "DOE-ER-40561-056, INT-92-07-01",
    doi = "10.1016/0370-2693(92)91887-F",
    journal = "Phys. Lett. B",
    volume = "287",
    pages = "128--132",
    year = "1992"
}

@article{Duan:2009cd,
    author = "Duan, Huaiyu and Kneller, James P.",
    title = "{Neutrino flavour transformation in supernovae}",
    eprint = "0904.0974",
    archivePrefix = "arXiv",
    primaryClass = "astro-ph.HE",
    reportNumber = "INT-PUB-08-27",
    doi = "10.1088/0954-3899/36/11/113201",
    journal = "J. Phys. G",
    volume = "36",
    pages = "113201",
    year = "2009"
}

@article{Duan:2010bg,
    author = "Duan, Huaiyu and Fuller, George M. and Qian, Yong-Zhong",
    title = "{Collective Neutrino Oscillations}",
    eprint = "1001.2799",
    archivePrefix = "arXiv",
    primaryClass = "hep-ph",
    reportNumber = "LA-UR-09-08309, INT-PUB-10-001",
    doi = "10.1146/annurev.nucl.012809.104524",
    journal = "Ann. Rev. Nucl. Part. Sci.",
    volume = "60",
    pages = "569--594",
    year = "2010"
}

@article{Mirizzi:2015eza,
    author = "Mirizzi, Alessandro and Tamborra, Irene and Janka, Hans-Thomas and Saviano, Ninetta and Scholberg, Kate and Bollig, Robert and H{\"u}depohl, Lorenz and Chakraborty, Sovan",
    title = "{Supernova Neutrinos: Production, Oscillations and Detection}",
    eprint = "1508.00785",
    archivePrefix = "arXiv",
    primaryClass = "astro-ph.HE",
    doi = "10.1393/ncr/i2016-10120-8",
    journal = "Riv. Nuovo Cim.",
    volume = "39",
    number = "1-2",
    pages = "1--112",
    year = "2016"
}

@article{Tamborra:2020cul,
    author = "Tamborra, Irene and Shalgar, Shashank",
    title = "{New Developments in Flavor Evolution of a Dense Neutrino Gas}",
    eprint = "2011.01948",
    archivePrefix = "arXiv",
    primaryClass = "astro-ph.HE",
    doi = "10.1146/annurev-nucl-102920-050505",
    journal = "Ann. Rev. Nucl. Part. Sci.",
    volume = "71",
    pages = "165--188",
    year = "2021"
}

@article{Padilla-Gay:2021haz,
    author = "Padilla-Gay, Ian and Tamborra, Irene and Raffelt, Georg G.",
    title = "{Neutrino Flavor Pendulum Reloaded: The Case of Fast Pairwise Conversion}",
    eprint = "2109.14627",
    archivePrefix = "arXiv",
    primaryClass = "astro-ph.HE",
    doi = "10.1103/PhysRevLett.128.121102",
    journal = "Phys. Rev. Lett.",
    volume = "128",
    number = "12",
    pages = "121102",
    year = "2022"
}

@ARTICLE{Yuzbashyan:2005-PRB,
       author = {{Yuzbashyan}, Emil A. and {Altshuler}, Boris L. and {Kuznetsov}, Vadim B. and {Enolskii}, Victor Z.},
        title = "{Nonequilibrium cooper pairing in the nonadiabatic regime}",
      journal = {Phys. Rev. B},
         year = 2005,
        month = dec,
       volume = {72},
       number = {22},
          eid = {220503},
        pages = {220503},
          doi = {10.1103/PhysRevB.72.220503},
archivePrefix = {arXiv},
       eprint = {cond-mat/0505493},
 primaryClass = {cond-mat.supr-con},
       adsurl = {https://ui.adsabs.harvard.edu/abs/2005PhRvB..72v0503Y}
}

@article{yuzbashyan2008normal,
  title={Normal and anomalous solitons in the theory of dynamical Cooper pairing},
  author={Yuzbashyan, Emil A},
  journal={Phys. Rev. B},
  volume={78},
  number={18},
  pages={184507},
  year={2008},
  eprint = "0807.3181",
  archivePrefix = "arXiv",
  doi = "10.1103/PhysRevB.78.184507",
  publisher={APS}
}

@article{yuzbashyan2005prb,
  title="{Integrable dynamics of coupled Fermi-Bose condensates}",
  author={Yuzbashyan, Emil A. and Kuznetsov, Vadim B. and Altshuler, Boris L.},
  journal={Phys. Rev. B},
  volume={72},
  pages={144524},
  year={2005},
  eprint = "cond-mat/0506782",
  archivePrefix = "arXiv",
  doi = "10.1103/PhysRevB.72.144524"
}

@article{yuzbashyan2005solution,
  title="{Solution for the dynamics of the BCS and central spin problems}",
  author={Yuzbashyan, Emil A. and Altshuler, Boris L. and Kuznetsov, Vadim B. and Enolskii, Victor Z.},
  journal={J. Phys. A: Math. Gen.},
  volume={38},
  number={36},
  pages={7831},
  year={2005},
  eprint = "cond-mat/0407501",
  archivePrefix = "arXiv",
  doi = "10.1088/0305-4470/38/36/003",
  publisher={IOP Publishing}
}

@ARTICLE{yuzbashyan:2006,
       author = {{Yuzbashyan}, Emil A. and {Tsyplyatyev}, Oleksandr and {Altshuler}, Boris L.},
        title = "{Relaxation and Persistent Oscillations of the Order Parameter in Fermionic Condensates}",
      journal = {Phys. Rev. Lett.},
         year = 2006,
        month = may,
       volume = {96},
        pages = {097005},
          doi = {10.1103/PhysRevLett.96.097005},
archivePrefix = {arXiv},
       eprint = {cond-mat/0511621},
       note  = "Erratum \href{https://doi.org/10.1103/PhysRevLett.96.179905}{{\em Phys. Rev. Lett.} {\bf 96} (2006) 179905}"
}

@article{Yuzbashyan:2018gbu,
    author = "Yuzbashyan, Emil A.",
    title = "{Integrable time-dependent Hamiltonians, solvable Landau\textendash{}Zener models and Gaudin magnets}",
    eprint = "1802.01571",
    archivePrefix = "arXiv",
    primaryClass = "cond-mat.stat-mech",
    doi = "10.1016/j.aop.2018.01.017",
    journal = "Annals Phys.",
    volume = "392",
    pages = "323--339",
    year = "2018"
}

@article{kuznetsov1992quadrics,
  title="{Quadrics on real Riemannian spaces of constant curvature: Separation of variables and connection with Gaudin magnet}",
  author={Kuznetsov, Vadim B},
  journal={J. Math. Phys.},
  volume={33},
  number={9},
  pages={3240--3254},
  year={1992},
  doi = "10.1063/1.529542",
  publisher={American Institute of Physics}
}

@article{Pehlivan:2011hp,
    author = "Pehlivan, Y. and Balantekin, A. B. and Kajino, Toshitaka and Yoshida, Takashi",
    title = "{Invariants of collective neutrino oscillations}",
    eprint = "1105.1182",
    archivePrefix = "arXiv",
    primaryClass = "astro-ph.CO",
    doi = "10.1103/PhysRevD.84.065008",
    journal = "Phys. Rev. D",
    volume = "84",
    pages = "065008",
    year = "2011"
}

@article{Hannestad:2006nj,
    author = "Hannestad, Steen and Raffelt, Georg G. and Sigl, Gunter and Wong, Yvonne Y. Y.",
    title = "{Self-induced conversion in dense neutrino gases: Pendulum in flavour space}",
    eprint = "astro-ph/0608695",
    archivePrefix = "arXiv",
    reportNumber = "MPP-2006-102",
    doi = "10.1103/PhysRevD.74.105010",
    journal = "Phys. Rev. D",
    volume = "74",
    pages = "105010",
    year = "2006",
    note = "Erratum: \href{https://doi.org/10.1103/PhysRevD.76.029901}{{\em Phys. Rev. D} {\bf 76} (2007) 029901}"
}

@article{Raffelt:2011yb,
    author = "Raffelt, Georg G.",
    title = "{N-mode coherence in collective neutrino oscillations}",
    eprint = "1103.2891",
    archivePrefix = "arXiv",
    primaryClass = "hep-ph",
    reportNumber = "MPP-2011-17, MPP-2011-17",
    doi = "10.1103/PhysRevD.83.105022",
    journal = "Phys. Rev. D",
    volume = "83",
    pages = "105022",
    year = "2011",
    note = "Erratum: \href{https://doi.org/10.1103/PhysRevD.104.089902}{{\em Phys. Rev. D} {\bf 104} (2021) 089902}"
}

@article{Johns:2019izj,
    author = "Johns, Lucas and Nagakura, Hiroki and Fuller, George M. and Burrows, Adam",
    title = "{Neutrino oscillations in supernovae: angular moments and fast instabilities}",
    eprint = "1910.05682",
    archivePrefix = "arXiv",
    NoprimaryClass = "hep-ph",
    doi = "10.1103/PhysRevD.101.043009",
    journal = "Phys. Rev. D",
    volume = "101",
    number = "4",
    pages = "043009",
    year = "2020"
}

@article{Duan:2007mv,
    author = "Duan, Huaiyu and Fuller, George M. and Carlson, J. and Qian, Yong-Zhong",
    title = "{Analysis of collective neutrino flavor transformation in supernovae}",
    eprint = "astro-ph/0703776",
    archivePrefix = "arXiv",
    doi = "10.1103/PhysRevD.75.125005",
    journal = "Phys. Rev. D",
    volume = "75",
    pages = "125005",
    year = "2007"
}

@article{Duan:2005cp,
    author = "Duan, Huaiyu and Fuller, George M. and Qian, Yong-Zhong",
    title = "{Collective neutrino flavor transformation in supernovae}",
    eprint = "astro-ph/0511275",
    archivePrefix = "arXiv",
    doi = "10.1103/PhysRevD.74.123004",
    journal = "Phys. Rev. D",
    volume = "74",
    pages = "123004",
    year = "2006"
}

@article{Gaudin:1976sv,
    author = "Gaudin, M.",
    title = "{Diagonalisation d'une classe d'Hamiltniens de spin}",
    reportNumber = "SACLAY-DPh-T/76-38",
    journal="J. Phys. (Paris)",
    volume = "37",
    pages="1087--1098",
    year = "1976"
}

\end{document}